\documentclass[12pt,a4paper]{article}
\pdfoutput=1
\usepackage[english]{babel} 
\usepackage{wrapfig}
\usepackage{epstopdf}
\usepackage{graphicx}
\usepackage{colordvi}
\usepackage{color}
\usepackage{latexsym}
\usepackage{amsmath}
\usepackage{amsfonts}
\usepackage{multirow}
\usepackage{bm}

\begin{document}
\newcommand{\nc}{\newcommand}
\nc{\beq}{\begin{equation}} \nc{\eeq}{\end{equation}}
\nc{\beqa}{\begin{eqnarray}} \nc{\eeqa}{\end{eqnarray}}
\nc{\eps}{\epsilon} \definecolor{Gray}{gray}{0.9}
\nc{\R}{{\cal R}}

\begin{center}
\vspace{1cm}

{\bf \large DIVERGENCES IN MAXIMAL SUPERSYMMETRIC\\[0.35CM] YANG-MILLS THEORIES\\[0.4CM] IN DIVERSE DIMENSIONS} \vspace{1.5cm}

{\bf \large L.V. Bork$^{2,4}$, D.I. Kazakov$^{1,2,3}$, M.V. Kompaniets$^5$,\\[0.4cm] D.M. Tolkachev$^{7,1}$ 
and D.E. Vlasenko$^{6,1}$}\vspace{0.7cm}

{\it $^1$Bogoliubov Laboratory of Theoretical Physics, Joint
Institute for Nuclear Research, Dubna, Russia.\\
$^2$Alikhanov Institute for Theoretical and Experimental Physics, Moscow, Russia\\
$^3$Moscow Institute of Physics and Technology, Dolgoprudny, Russia\\
$^4$Center for Fundamental and Applied Research, All-Russian Institute of Automatics, Moscow, Russia \\ $^5$St. Petersburg State University, St. Petersburg, Russia\\
$^6$Department of Physics, Southern Federal State University, Rostov-Don, Russia\\ and \\
$^7$Department of Physics, Gomel State University, Gomel, Belarus}
\vspace{0.5cm}

\abstract{The main aim of this paper  is to study the scattering amplitudes in gauge field  theories with maximal supersymmetry in dimensions $D=6,8$ and $10$.
We perform a systematic study  of the leading ultraviolet divergences using the spinor helicity and on-shell momentum superspace framework. In $D=6$ the first divergences start at 3 loops and we calculate them up to 5 loops, in $D=8,10$  the first divergences start at 1 loop and we calculate them up to 4 loops. The leading divergences in a given order are the polynomials of Mandelstam variables. To be on the safe side, we check our analytical calculations by numerical ones applying the alpha-representation and  the dedicated routines. Then we derive an analog of the RG equations for the leading pole
that allows us to get the recursive relations  and construct the generating procedure to obtain the  polynomials at any order of perturbation theory (PT). At last, we make an attempt to sum the PT series and derive the differential equation for the infinite sum. This equation possesses a fixed point which might be stable or unstable depending on the kinematics. Some consequences of these fixed points are discussed. }
\end{center}


\newpage

\tableofcontents{}
\vspace{0.5cm}
{\bf References\hfill 40}
\newpage

\section{Introduction}\label{Introduction_1}
In the recent decade there was considerable progress in understanding the structure of  the amplitudes (the S-matrix) in gauge theories in various dimensions (for review see, for example,
\cite{Reviews_Ampl_General}).   The gauge and gravity theories with maximal supersymmetries in $D=4$, $\mathcal{N}=4$ SYM and $\mathcal{N}=8$ SUGRA are the most important examples. This progress became possible due to the development of the new  techniques: the spinor helicity and momentum twistor formalisms, different sets of recurrence relations for the tree level amplitudes, the unitarity  based methods for the loop amplitudes and various realizations of the on-shell superspace formalism for
theories with supersymmetry \cite{Reviews_Ampl_General}. 

The subject of investigation was mainly related to the so-called maximally supersym\-metric theories, which are believed to possess special properties due the highest symmetries. One of the insights  was the discovery of the dual conformal symmetry for the $\mathcal{N}=4$ SYM. Taking together the algebras of ordinary (super)conformal symmetry and the dual (super)conformal symmetry can be fused into an
infinite dimensional Yangian algebra \cite{Drummond_YangianSymmetryAmplitudes} which in principle should
completely define the $S$-matrix of the $\mathcal{N}=4$ SYM theory
\cite{BeisertYangianRev,StaudacherNew,Viera}.

While the $\mathcal{N}=4$ SYM theory is completely  on shell  UV finite and possesses  only the IR divergences, in higher dimensions the situation is the opposite: there are no IR divergences even on shell but all theories are UV nonrenormalizable by power counting. 

It should be noted that the spinor helicity formalism and  the unitarity based methods
can be generalised to space-time dimension greater than $D=4$
\cite{SpinorHelisityForm_D=10Dimentions, SpinorHelisityForm_GeneralDimentions_Boels,DonaldOConnel_AmplInD=6}.
The new computational methods gave new birth to the investigation of the UV properties
of the S-matrices of formally nonrenormalizable gravity theories with extended
supersymmetry ($D=4$ $\mathcal{N}=8$ SUGRA is a particular example). The results obtained so far are in some sense controversial \cite{N=8SUGRA finiteness,SUGRA fin Vs Div,D=5SYM_Diverges_ZBernDixon}. 

Among the gauge theories in higher dimensions with maximal supersymmetry there are the following four cases:
$$ D=4\  \mathcal{N}=4, \ \ D=6 \ \mathcal{N}=2, \ \  D=8 \ \mathcal{N}=1, \ \ D=10 \ \mathcal{N}=1.$$
No wonder if all these theories obey some exceptional properties. 
In this context, it is interesting to note that the  integrands of the four-point amplitudes in any SYM theory have almost identical form (only the tree level amplitudes which are the common factors are different) and are heavily constrained by the dual conformal covariance in dimensions $D\leq6$ \cite{D6_DualConformal_Invariance} (and likely in all dimensions $D\leq 10$ \cite{SpinorHelisityForm_D=10Dimentions}).

The aim of this paper, which is a continuation of our previous papers \cite{BKV,BKV1K}, is to investigate  the amplitudes and their UV properties in maximally supersymmetric gauge theories in various dimensions.
Namely, we evaluate the leading UV divergences in the four-point amplitude on shell in a number of loops and investigate their properties in all loops. The paper is organized as follows: in section 2, we briefly describe the  spinor helicity formalism in $D\geq 6$, and in section 3, we consider the on shell superspace formalism in $D=6,8$. In section 4, we consider the structure of the colour ordered  partial  amplitudes.
In section 5, we present the evaluation  of the leading UV divergences  both analytically and numerically.   Section 6  summarizes   the results  of  perturbative computation for various dimensions. Finally in section 7 we derive the all loop recursive relations for the leading divergences and make the attempts to summarize the whole PT series. The conclusion contains  some speculations regarding the observed pattern and its implications on possible scenarios of UV finiteness of gauge and gravity theories. 

\section{The spinor helicity formalism in various dimensions}\label{_21}

As was stated in the introduction, the spinor helicity and the on shell momentum superspace formalisms  play a crucial role in the recent achievements in understanding the structure of the S-matrix of four dimensional supersymmetric gauge field theories. Here we discuss the generalization of these formalism to the case of  even dimensions $D=6,8$ and $10$. 

In our discussion we manly follow \cite{Boles_Spinors_D8}. In even dimensions one can always choose the chiral representation of the gamma matrices as $\Gamma^{\mu}$ (as usual $\{\Gamma^{\mu},\Gamma^{\nu}\}=2\eta^{\mu\nu}$):
\begin{equation}
 \Gamma^{\mu}=\left(
\begin{array}{ccc}
  0 & (\sigma^{\mu})^{AB'} \\
 (\overline{\sigma}^{\mu})_{B'A} & 0 \\
 \end{array}
\right),
\end{equation}
where $\mu$ is the $SO(D-1,1)$ vector representation index, $A$ and $B'=1,...,2^{D/2-1}$ are the $Spin(SO(D-1,1))$ indices. We are interested in the cases $D=4,6,8,10$. In this notation one can decompose the Dirac spinor $\psi$ as a pair of Weyl chiral and anti-chiral spinors $\lambda^A$ and $\tilde{\lambda}_{A'}$.
The Lorentz rotations of $\lambda^A$ and $\tilde{\lambda}_{A'}$ look like
\begin{eqnarray}
\delta\lambda^A=(\sigma^{\mu\nu})^A_B\lambda^B,~\delta\tilde{\lambda}_{A'}=(\sigma^{\mu\nu})^{B'}_{A'}\tilde{\lambda}_{B'},
\end{eqnarray}
where
\begin{eqnarray}
(\sigma^{\mu\nu})^A_B&\equiv&
\frac{i}{4}\left[(\sigma^{\mu})^{AA'}(\overline{\sigma}^{\nu})_{A'B}-
(\sigma^{\nu})^{AA'}(\overline{\sigma}^{\mu})_{A'B}\right]\nonumber\\
(\sigma^{\mu\nu})^{B'}_{A'}&\equiv&\frac{i}{4}
\left[(\overline{\sigma}^{\mu})_{A'A}(\sigma^{\nu})^{AB'}-
(\overline{\sigma}^{\nu})^{A'A}(\sigma^{\mu})_{AB'}\right].
\end{eqnarray}
One can combine two Dirac spinors $\psi_2$ and $\psi_1$ into the Lorentz invariant combination
$\psi_1^TC\psi_2$
using the charge conjugation matrix $C$ defined so that
\begin{eqnarray}
C\Gamma^{\mu}C^{-1}=-(\Gamma^{\mu})^T.
\end{eqnarray}
For the Weil spinors there are two possible decompositions of $C$ depending on dimension:
\begin{equation}
 C=\left(
\begin{array}{ccc}
  \Omega_{BA} & 0 \\
 0 & \Omega^{B'A'} \\
 \end{array}
\right)~\mbox{for}~D=4,8,
\end{equation}
and
\begin{equation}\label{OmegaD610}
 C=\left(
\begin{array}{ccc}
  \Omega_{B}^{A'} & 0 \\
 0 & \Omega^{B'}_{A} \\
 \end{array}
\right)\mbox{for}~D=6,10.
\end{equation}
The $\Omega$ matrices obey the following relations
\begin{equation}
  \Omega_{BA} \Omega^{AC}=\delta^C_B,~ \Omega_{B'A'} \Omega^{A'C'}=\delta^{B'}_{C'}~\mbox{for}~D=4,8,
\end{equation}
and
\begin{equation}
\Omega_B^{A'}\Omega_{A'}^C=\delta^C_B,~\Omega_A^{B'}\Omega_{C'}^A=\delta^{B'}_{C'}~\mbox{for}~D=6,10.
\end{equation}
The matrices $\Omega$ can be used to raise and lower the indices of the spinors
\begin{eqnarray}
\lambda_A=\lambda^B\Omega_{BA}~\tilde{\lambda}^{A'}=\Omega^{B'A'}\tilde{\lambda}_{B'}~\mbox{for}~D=4,8,
\end{eqnarray}
and to relate the chiral and antichiral spinors
\begin{eqnarray}
\lambda_A=\Omega^{A'}_A\tilde{\lambda}_{A'},~\tilde{\lambda}^{A'}=\lambda^A\Omega_A^{A'}~\mbox{for}~D=6,10.
\end{eqnarray}
One can also construct the Lorentz invariants for the pair of spinors which are labeled  by $i$ and $j$:
\begin{eqnarray}
\lambda^B_i\Omega_{BA}\lambda^A_j\equiv \langle ij \rangle,
~\tilde{\lambda}_{B',i}\Omega^{B'A'}\tilde{\lambda}_{A',j}\equiv[ ij ]~\mbox{for}~D=4,8,
\end{eqnarray}
and
\begin{eqnarray}
\tilde{\lambda}_{A',i}\Omega^{A'}_A\lambda^A_j\equiv[i|j\rangle,
~\lambda^A_i\Omega_A^{A'}\lambda_{A',j}\equiv\langle i|j]~\mbox{for}~D=6,10.
\end{eqnarray}
The matrices $C$ can be always chosen in such a way that
\begin{eqnarray}
C^T&=&-C~\mbox{for}~D=4,10,\nonumber\\
C^T&=&C~\mbox{for}~D=6,8.
\end{eqnarray}
In some dimensions one can also construct additional Lorentz invariants. For example, in $D=6$ one has
$Spin(SO(5,1))\cong SU(4)^*$, so one can construct the invariants as contractions of spinorial indices $A$ with the  absolutely antisymmetric tensor $\varepsilon_{ABCE}$ asso\-siated with $SU(4)^*$:
$\epsilon_{ABCD}\lambda^{A}_1\lambda^{B}_2\lambda^{C}_3\lambda^{D}_4
  \equiv \langle 1234 \rangle$, $\epsilon^{ABCD}\tilde{\lambda}_{A,1}\tilde{\lambda}_{B,2}\tilde{\lambda}_{C,3}\tilde{\lambda}_{D,4}\equiv [ 1234 ]$.

To relate the light like (massless) momentum $p_{\mu}$ with the pair of Weyl spinors we consider the  Dirac equations for the spinors $\lambda^A$ and $\tilde{\lambda}_{A'}$:
\begin{eqnarray}
  (p_{\mu}\sigma^{\mu})^{BA'}\tilde{\lambda}_{A'}=0~\mbox{and}~
  (p_{\mu}\tilde{\sigma}^{\mu})\lambda^A=0.
\end{eqnarray}
The solutions to these equations are labeled by additional helicity indices
$a$ and $a'$ which transform under the little group of the Lorentz group, which is
$SO(D-2)$ in our  case.  Note that in $D > 4$ dimensions helicity of a massless particle is no longer conserved and transforms according to the little group similarly  to helicity of a massive particle in $D=4$.
One has
\begin{eqnarray}
  (p_{\mu}\sigma^{\mu})^{BA'}\tilde{\lambda}_{A'a'}=0,~
  (p_{\mu}\tilde{\sigma}^{\mu})\lambda^{Aa}=0,
\end{eqnarray}
and for their conjugates
\begin{eqnarray}
  (p_{\mu}\sigma^{\mu})^{BA'}\lambda_{B}^{a'}=0,~
  (p_{\mu}\tilde{\sigma}^{\mu})\tilde{\lambda}^{A'}_a=0.
\end{eqnarray}
It is always possible to take the solutions to these equations
$\tilde{\lambda}_{A'a'}(p),\lambda^{Aa}(p)$ (and their conjugates) in
such a way that
\begin{eqnarray}\label{SpinorToMomentum}
  \sum_{a}\lambda^{Ba}(p)\tilde{\lambda}^{A'}_{a}(p)=p_{\mu}(\sigma^{\mu})^{BA'},~
  \sum_{a'}\tilde{\lambda}_{B'a'}(p)\lambda^{a'}_{A}(p)=p_{\mu}(\tilde{\sigma}^{\mu})_{B'A}.
\end{eqnarray}
One can choose the polarization vectors of gluons in the form
\begin{eqnarray}
  \varepsilon^{\mu}_{aa'}(p|q)\sim q_{\nu}\frac{\tilde{\lambda}_{a}(p)(\overline{\sigma}^{\mu}\sigma^{\nu})\tilde{\lambda}_{a'}(q)}{(pq)},~
  \varepsilon^{\mu,aa'}(p|q)\sim q_{\nu}\frac{\lambda^{a}(p)(\sigma^{\mu}\sigma^{\nu})\lambda^{a'}(q)}{(pq)}.
\end{eqnarray}
The polarization vectors for massless fermions can also be  chosen as spinors while the polarization vectors for scalars are trivial.
The dependence of the polarization vectors on the auxilary light like momenta $q$ reflects the gauge ambiguity in the choice of polarization vectors. The dependence on $q$ is always canceled in the final gauge invariant amplitude.

This way one can always write down the scattering amplitude in the gauge theory in arbitrary even dimension as a function of the  Lorentz invariant products of  momenta and polarization  vectors  in terms of the spinor products.

Concluding this section we would like to comment on the difference between the spinor representation of the amplitudes in $D=4,6$ and $D=8,10$ dimensions.
In $D=4$ the little group is $SO(2)\simeq U(1)$, so its action on the  spinors is just a multiplication by a complex number. 
The condition $p^2=0$ for $p^{\alpha\dot{\alpha}}=p_{\mu}(\sigma^{\mu})^{\alpha\dot{\alpha}}$
is equivalent to $det(p)=0$, so the following equality holds:
\begin{eqnarray}
p^{\alpha\dot{\alpha}}=\lambda^{\alpha}\tilde{\lambda}^{\dot{\alpha}},
\label{spinmom}.
\end{eqnarray}
On the other hand, one can always use the solutions of the Dirac equation $\lambda^{\alpha}(p)$, $\tilde{\lambda}^{\dot{\alpha}}(p)$ to write
\begin{eqnarray}
\lambda^{\alpha}(p)\tilde{\lambda}^{\dot{\alpha}}(p)=p^{\alpha\dot{\alpha}}.
\end{eqnarray}
This means that relation (\ref{spinmom}) works in "both directions".
For given spinors $\lambda^{\alpha}$ and $\tilde{\lambda}^{\dot{\alpha}}$
there is always  (complex) momentum $p$ such that $p^{\alpha\dot{\alpha}}=\lambda^{\alpha}\tilde{\lambda}^{\dot{\alpha}}$
and vice versa one can always decompose the light-like momentum $p$ into a pair of spinors using the solution of the Dirac equation. This is possible because the product of two $D=4$ Weyl spinors  $\lambda^{\alpha}$ and $\tilde{\lambda}^{\dot{\alpha}}$
contains $2\times2-1=3$ ($-1$ is due to the little group $U(1)$ invariance of $\lambda^{\alpha}\tilde{\lambda}^{\dot{\alpha}}$)
independent components, just  as the light-like momentum $p_{\mu}$.

The same situation occurs in $D=6$. One has  $4\times2$ components in the spinor product, and taking into account the action of the $SU(2)$ little group ($SO(6-2)\simeq SU(2) \times SU(2)$) one gets
 $4\times2-3=5$ degrees of freedom, exactly as for the $D=6$ massless momentum $p_{\mu}$.

For $D>6$ this is no longer the case. While it is still possible for a given $p_{\mu}$ to find solutions of the Dirac equation such that (\ref{spinmom}) holds, for a given set of spinors in  $D>6$ there is no unique $p_{\mu}$ satisfying this equation \cite{SpinorHelisityForm_GeneralDimentions_Boels,SpinorHelisityForm_D=10Dimentions}. 
In other words, one may say that for $D>6$ the spinors obey the nonlinear relations (constraints)\cite{SpinorHelisityForm_D=10Dimentions} and one way to solve these constraints  is to require that they satisfy the Dirac equation for some  light like momenta $p_{\mu}$ \cite{SpinorHelisityForm_D=10Dimentions}. This means that the spinor helicity formalism in $D=8,10$ may not be  optimal (in terms of simplicity) representation or amplitudes.

\section{The on-shell momentum superspace in various \\ dimensions}

In the next sections, we  discuss the essential details regarding the
on shell momentum superspace
constructions in  $D=6,8,10$ dimensions.

\subsection{$D=6$ $\mathcal{N}=(1,1)$ SYM}\label{_22}

The usage of the  on shell momentum superspace allows one to  obtain a compact represen\-tation for
the amplitudes in supersymmetric gauge theories, 
which is very  convenient in the unitarity based computations \cite{Reviews_Ampl_General}.

Consider now the essential part of the $D=6$ $\mathcal{N}=(1,1)$
on-shell momentum superspace formalism.
The on-shell
$\mathcal{N}=(1,1)$  superspace for $D=6$ SYM was first formulated in \cite{D6_DualConformal_Invariance,ZBern_GenUnit_D=6Helicity,Sigel_D=6Formalism}.
It can be parameterized by the following set of coordinates:
\begin{eqnarray}\label{Full_(1,1)_superspace}
  \mbox{$\mathcal{N}=(1,1)$ D=6 on-shell superspace}=\{\lambda^A_a,\tilde{\lambda}_{A}^{\dot{a}},\eta_a^I,\overline{\eta}_{I'\dot{a}}\},
\end{eqnarray}
where $\eta_a^I$ and $\overline{\eta}_{\dot{a}}^{I'}$ are the Grassmannian coordinates,
$I=1,2$ and $I'=1',2'$ are the $SU(2)_R\times SU(2)_R$ R-symmetry indices. Note that
this superspace is not chiral. One has two types of the supercharges $q^{A I}$ and
$\overline{q}_{A I'}$ with the commutation relations
\begin{eqnarray}\label{commutators_for_superchrges_full_(1,1)}
  \{ q^{A I}, q^{B J}\}&=&p^{AB}\epsilon^{IJ},\nonumber\\
  \{ \overline{q}_{A I'}, \overline{q}_{B J'}\}&=&p_{AB}\epsilon_{I'J'},\nonumber\\
  \{ q^{A I},  \overline{q}_{B J'}\}&=&0.
\end{eqnarray}

The creation/annihilation operators of the on shell states from the  $\mathcal{N}=(1,1)$ supermultiplet are
$$
\{A_{a\dot{a}},~\Psi^a_I,~\overline{\Psi}^{I'\dot{a}},~\phi^{I'}_I\},
$$
which corresponds to the physical polarizations of the gluon $|A_{a\dot{a}}\rangle$, two fermions
$|\Psi^a_I\rangle$,$|\overline{\Psi}^{I'\dot{a}}\rangle$ and two complex scalars
$|\phi^{I'}_I\rangle$ (antisymmetric with respect to $I,I'$). This multiplet is CPT
self-conjugated. However, to combine all the on-shell states in one
superstate $|\Omega\rangle$ by analogy with the $\mathcal{N}=4$
$D=4$ SYM \cite{DualConfInvForAmplitudesCorch}, 
one has to perform a truncation of the full
$\mathcal{N}=(1,1)$ on-shell superspace \cite{Sigel_D=6Formalism} in
contrast to the former case. Indeed, if one expands any function $X$
(or $|\Omega\rangle$ superstate) defined on the full on-shell
superspace in Grassmannian variables, one encounters  terms like
$\sim\eta_a^I\overline{\eta}_{I'\dot{a}}A_{I}^{I'a\dot{a}}$. Since
there are no such bosonic states $A_{I}^{I'a\dot{a}}$ in the
$\mathcal{N}=(1,1)$ SYM supermultiplet,  one needs to eliminate these
terms by imposing constraints on $X$, i.e.,  to truncate the full
on-shell superspace. If one wishes to use the little group indices
to label the on-shell states, the truncation has to be done with
respect to the R symmetry indices. This can be done by consistently
using the harmonic superspace techniques \cite{Sigel_D=6Formalism}.

Defining the harmonic variables $u_{I}^{\mp}$ and $\overline{u}^{\pm
I'}$ which parameterize the double coset space
\begin{eqnarray}
\frac{SU(2)_R}{U(1)}\times\frac{SU(2)_R}{U(1)}
\end{eqnarray}
we express  the projected supercharges, the Grassmannian coordinates
\begin{eqnarray}
  q^{\mp A}&=&u^{\mp}_Iq^{A I},~
  \overline{q}^{\pm}_{A}=u^{\pm I'}\overline{q}_{A I'},
  \nonumber\\
  \eta^{\mp}_a&=&u^{\mp}_I\eta_a^I,~
  \overline{\eta}^{\pm}_{\dot{a}}=u^{\pm I'}\overline{\eta}_{I'\dot{a}},
\end{eqnarray}
and the creation/annihilation operators of the on-shell states
\begin{eqnarray}\label{(1,1)_onshel_states}
  &&\phi^{--},~\phi^{-+},~\phi^{+-},~\phi^{++},\nonumber\\
  &&\Psi^{-a},~\Psi^{+a},~\overline{\Psi}^{-\dot{a}}~\overline{\Psi}^{+\dot{a}},\nonumber\\
  &&A^{a\dot{a}}.
\end{eqnarray}
in terms of the new harmonic variables.

In  what follows we consider only the objects that depend on the set of
variables which parameterize the subspace ("analytic superspace") of
the full $\mathcal{N}=(1,1)$ on-shell superspace
\begin{eqnarray}\label{Truncated_onshell_superspace}
  \mbox{$\mathcal{N}=(1,1)$ D=6 on-shell harmonic  superspace}=\{\lambda^A_a,\tilde{\lambda}_{A}^{\dot{a}},
\eta^{-}_a,\overline{\eta}_{\dot{a}}^{+} \}.
\end{eqnarray}
The projected supercharges and momentum generators acting on the analytic superspace
for the n-particle case can be explicitly written as:
\begin{eqnarray}\label{projected_supercharges_n_particle_state}
  p^{AB}=\sum_i^n\lambda^{Aa}(i)\lambda^B_{a}(i),~ q^{-A}=\sum_i^n\lambda^A_a(i)\eta^{-a}_i,~
  \overline{q}_A^+=\sum_i^n\tilde{\lambda}_A^{\dot{a}}(i)\overline{\eta}_{\dot{a},i}^+.
\end{eqnarray}
Now one can combine all the on-shell state creation/annihilation operators
(\ref{(1,1)_onshel_states}) into
one superstate $|\Omega_i\rangle=\Omega_i|0\rangle$ (here $i$ labels the momenta
carried by the state):
\begin{eqnarray}
  |\Omega_i\rangle&=&\{ \phi^{-+}_i+\phi^{++}_i(\eta^-\eta^-)_i+
  \phi^{--}_i(\overline{\eta}^+\overline{\eta}^+)_i
  +\phi^{+-}_i(\eta^-\eta^-)_i(\overline{\eta}^+\overline{\eta}^+)_i
  \nonumber\\
  &+&(\Psi^+\eta^-)_i+(\overline{\Psi}^-\overline{\eta}^+)_i+
  (\Psi^-\eta^-)_i(\overline{\eta}^+\overline{\eta}^+)_i+
  (\overline{\Psi}^+\overline{\eta}^+)_i(\eta^-\eta^-)_i\nonumber\\
  &+&(A\eta^-\overline{\eta}^+)_i\}|0\rangle,
\end{eqnarray}
where $(XY)_i\doteq X^{a/\dot{a}}_iY_{i~a/\dot{a}}$. Hereafter we
will drop the $\pm$ labels for simplicity.
As in $D=4$ case, we can formally write the colour ordered amplitude as
\begin{eqnarray}
 \mathcal{A}_n(\{\lambda^A_a,\tilde{\lambda}_{A}^{\dot{a}},
\eta_a,\overline{\eta}_{\dot{a}} \})=\langle
0|\prod_{i=1}^n\Omega_i S|0\rangle,
\end{eqnarray}
Here $S$ is the S-matix operator of the theory, the average $\langle0| \ldots |0\rangle$ is understood with
respect to some particular (for example, component) formulation of
the theory.
The invariance with respect to translations and supersymmetry transformations
requires that
\begin{eqnarray}
 p^{AB}\mathcal{A}_n=q^A\mathcal{A}_n=\overline{q}_A\mathcal{A}_n=0.
\end{eqnarray}
Thus,  the  superamplitude should have the form:
\begin{eqnarray}
 \mathcal{A}_n(\{\lambda^A_a,\tilde{\lambda}_{A}^{\dot{a}},
\eta_a,\overline{\eta}_{\dot{a}} \})=
\delta^6(p^{AB})\delta^4(q^A)\delta^4(\overline{q}_A)\mathcal{P}_n(\{\lambda^A_a,\tilde{\lambda}_{A}^{\dot{a}},
\eta_a,\overline{\eta}_{\dot{a}} \}),
\end{eqnarray}
where $\mathcal{P}_n$ is a polynomial with respect to
$\eta$ and $\overline{\eta}$ of degree  $2n-8$. Note that since there is no helicity as a conserved quantum number,
there are no closed subsets of MHV, NMHV, etc. amplitudes in contrast to the $D=4$
case.  

The Grassmannian  delta functions $\delta^4(q^A)$ and 
$\delta^4(\overline{q}_A)$ are defined
in this case as
\begin{eqnarray}
  \delta^4(q^A)&=&\frac{1}{4!}\epsilon_{ABCD}
  \hat{\delta}(q^A)\hat{\delta}(q^B)
  \hat{\delta}(q^C)\hat{\delta}(q^D),\nonumber\\
  \delta^4(\overline{q}_A)&=&\frac{1}{4!}\epsilon^{ABCD}
  \hat{\delta}(\overline{q}_{A})\hat{\delta}(\overline{q}_{B})
  \hat{\delta}(\overline{q}_{C})\hat{\delta}(\overline{q}_{D}).
\end{eqnarray}
Here the delta function $\hat{\delta}(X^I)$ is the usual  Grassmannian delta function defined as $\hat{\delta}^N(X^I) \equiv \prod_{I=1}^N X^I$, where $I$ is the R-symmetry index. In harmonic formulation we simply have $\hat{\delta}(X) \equiv X$.

To extract the ordinary component amplitudes from this supersymmetric expression,
one has to apply the projection operators.  The projection operators are the derivatives with respect to an appropriate number of Grassmannian variables. Their explicit form
can be read from (\ref{superstate}). For instance, the projection operator for the i-th gluon is $\partial/\partial \eta_i^-\partial/\partial \bar{\eta}_i^+$.

Consider now the four-point amplitude.
The degree of  the Grassmannian polynomial $\mathcal{P}_4$ is
$2n-8=0$, so $\mathcal{P}_4$ is a function of bosonic variables
$\{\lambda^A_a,\tilde{\lambda}_{A}^{\dot{a}}\}$ only, just as in the $D=4$ case
\begin{eqnarray}\label{4-point_ampl_general_form}
  \mathcal{A}_4(\{\lambda^A_a,\tilde{\lambda}_{A}^{\dot{a}},
\eta_a,\overline{\eta}_{\dot{a}} \})=
  \delta^6(p^{AB})\delta^4(q^A)\delta^4(\overline{q}_A)
  \mathcal{P}_4(\{\lambda^A_a,\tilde{\lambda}_{A}^{\dot{a}}\}).
\end{eqnarray}
At the tree level $\mathcal{P}_4$ can be found from the explicit expression  for the 4 gluon amplitude 
in components\cite{DonaldOConnel_AmplInD=6,Sigel_D=6Formalism}
obtained with the help of  the six dimensional version of the BCFW recurrence relation
\cite{DonaldOConnel_AmplInD=6,ZBern_GenUnit_D=6Helicity}.
Comparing the component expression with (\ref{4-point_ampl_general_form}) and
expanding (\ref{4-point_ampl_general_form}) in powers of $\eta$
one concludes that: $\mathcal{P}_4^{(0)}\sim1/st$,
where $s$ and $t$ are the standard Mandelstam
variables.
So at the tree level the 4-point
superamplitude can be written as:
\begin{eqnarray}\label{4_point_tree_superamplitude}
  \mathcal{A}_4^{(0)}=\delta^6(p^{AB})\delta^4(q^A)\delta^4(\overline{q}_A)\frac{1}{st}.
\end{eqnarray}
Note that already at the tree level the 5-point amplitude is not so simple
\cite{DonaldOConnel_AmplInD=6,Sigel_D=6Formalism}. Similarly to the $D=4$ case,
one can obtain the expression up to three loops using the  iterated two particle cuts.
To perform this computation, the following formula for the Grassmannian integration is useful:
($\int d^2 \eta^a_{l_1}~\int d^2\overline{\eta}^{\dot{b}}_{l_2}
  \equiv \int d^4\eta_{l_1l_2}$)
\begin{eqnarray}\label{Two_particle_Grassmann_int}
  &&\int d^4\eta_{l_1l_2}d^4\eta_{l_2l_1}~
  \delta^4(\lambda^{Aa}_{l_1} \eta_{a,l_1} +\lambda^{Aa}_{l_2} \eta_{a,l_2} + q^A_1)\delta^4(\lambda^{Aa}_{l_1} \eta_{a,l_1} +\lambda^{Aa}_{l_2} \eta_{a,l_2} \eta-q^A_2)
  \nonumber\\&&\times\delta^4(\tilde{\lambda}^{A\dot{a}}_{l_1}\overline{\eta}_{\dot{a},l_1}+\tilde{\lambda}^{A\dot{a}}_{l_2}\overline{\eta}_{\dot{a},l_2}+\overline{q}_{B})
  \delta^4(\tilde{\lambda}^{A\dot{a}}_{l_1}\overline{\eta}_{\dot{a},l_1}+\tilde{\lambda}^{A\dot{a}}_{l_2}\overline{\eta}_{\dot{a},l_2}-\overline{q}_{B})
  \nonumber\\&&=(2!)^44(l_1,l_2)^2\delta^4(q_1^A+q_2^A)\delta^4(\overline{q}_{B,1}+\overline{q}_{B,2}).
\end{eqnarray}

\subsection{$D=8$ $\mathcal{N}=1$ SYM}\label{_23}

Consider now the $D=8$ $\mathcal{N}=1$ case. The details of the on-shell
$\mathcal{N}=1$  superspace for $D=8$ SYM can be found  in \cite{Boles_Spinors_D8}.
It can be parameterized by the following set of coordinates
\begin{eqnarray}\label{Full_N=1 D=8_superspace}
  \mbox{$\mathcal{N}=1$ D=8 on-shell superspace}=\{\lambda^{Aa},\tilde{\lambda}_{a}^{A'},\eta_a\},
\end{eqnarray}
where $\eta^a$ are the Grassmannian coordinates,
$a$ is the  little group $SO(6)$ index, and $A$ and $A'$ are the $spin(SO(7,1))$ indices. The R-symmetry
group here is $U(1)_R$  and $\eta^a$ carries  the $+1$  charge with respect to $U(1)_R$. Note that
this superspace is chiral.

The commutation relations for the supercharges have the usual form
\begin{eqnarray}\label{commutators_for_superchrges_N=1 D=8}
  \{ q^{A}, \bar{q}^{B'}\}&=&p^{AB},
\end{eqnarray}
where the supercharges in the on-shell momentum superspace representation for the  $n$-particle case are
\begin{eqnarray}\label{commutators_for_superchrges_N=1 D=8 details}
   p^{AB'}=\sum_{i=1}^n\lambda^{Aa}(i)\tilde{\lambda}_a^{B'}(i),
   ~q^{A}=\sum_{i=1}^n\lambda^{Aa}(i)\eta_a, ~\bar{q}^{B'}=\sum_{i=1}^n\tilde{\lambda}_a^{B'}(i)\frac{\partial}{\partial\eta_a}.
\end{eqnarray}
The creation/annihilation operator states in the $\mathcal{N}=1$ $D=8 $ on-shell supermultiplet are
$$
\{A^{a\dot{a}},~\Psi^a,~\overline{\Psi}_{a},~\phi,~\overline{\phi}\},
$$
which corresponds to the  physical polarizations of the gluon $|A^{a\dot{a}}\rangle$,
two fermions $|\Psi^a\rangle$, $|\overline{\Psi}_{a}\rangle$ and two scalars $|\phi\rangle$, $|\overline{\phi}\rangle$.
One can combine them into one "superstate" $|\Omega_i\rangle$ similar to the $D=4$
case
\begin{eqnarray}\label{superstate}
|\Omega_{i}\rangle = \left(\phi_i + \eta_a\Psi^a_i +
\frac{1}{2!}\eta_a\eta_b A^{a\dot{a}}_i +
\frac{1}{3!}\eta_a\eta_b\eta_c \varepsilon^{abcd}\overline{\Psi}_{d,i} +
\frac{1}{4!}\eta_a\eta_b\eta_c\eta_d \varepsilon^{abcd}\overline{\phi}_i\right) |0\rangle.
\end{eqnarray}
Here $\varepsilon^{abcd}$ is the absolutely antisymmetric tensor associated with the little group $SO(6)\cong SU(4)$. Using the arguments identical to the $D=4,6$ cases we conclude that the colour ordered superamplitude should have the form:
\begin{eqnarray}
 \mathcal{A}_n(\{\lambda^{Aa},\tilde{\lambda}_{a}^{A'},\eta_a \})=
\delta^8(p^{AB'})\delta^8(q^A)\mathcal{P}_n(\{\lambda^{Aa},\tilde{\lambda}_{a}^{A'},\eta_a \}),
\end{eqnarray}
where $\mathcal{P}_n$ is a polynomial with respect to
$\eta$ and $\overline{\eta}$ of degree $2n-8$.
The Grassmannian  delta function $\delta^8(q^A)$ is defined
in this case as:
\begin{eqnarray}
  \delta^8(q^A)=\frac{1}{8!}\epsilon_{A_1...A_8}
  \prod_{i=1}^8\hat{\delta}(q^{A_i}),
\end{eqnarray}
Here $\epsilon_{A_1...A_8}$ is the 
absolutely antisymmetric tensor associated with the $spin(SO(7,1))$.
 
Consider the four-point  amplitude.
The degree of Grassmannian polynomial $\mathcal{P}_4$ is
$2n-8=0$, so as in the previous cases $\mathcal{P}_4$ is a function of bosonic variables
and one can  again write the four-point amplitude in the form 
\begin{eqnarray}\label{4-point_ampl_general_form8}
  \mathcal{A}_4(\{\lambda^{Aa},\tilde{\lambda}_{a}^{A'},\eta_a \})=
  \delta^8(p^{AB'})\delta^4(q^A)
  \mathcal{P}_4(\{\lambda^{Aa},\tilde{\lambda}_{a}^{A'}\}).
\end{eqnarray}
At the tree level $\mathcal{P}_4$ can be found from a comparison
with the explicit expression in the components obtained as the  field theory limit of the string
scattering amplitude.  As in the $D=6$ case, one has $\mathcal{P}_4^{(0)}\sim 1/st$.
So at the tree level the 4-point superamplitude can again be written as:
\begin{eqnarray}\label{4_point D8_tree_superamplitude}
  \mathcal{A}_4^{(0)}=\delta^8(p^{AB})\delta^8(q^A)\frac{1}{st}.
\end{eqnarray}
Similarly to the $D=4,6$ cases,
one can obtain the expression up to three loops using the iterated two particle cuts.
To perform this computation, the following formula for the Grassmannian integration is useful, which is similar to the $D=4$ case:
\begin{eqnarray}
  &&\int d^4\eta_{l_1}d^4\eta_{l_2}
  \delta^8(\lambda_{l_1}^{Aa}\eta_{a,l_1}+\lambda_{l_2}^{Aa}\eta_{a,l_2}+q_1^A)
  \delta^8(\lambda_{l_1}^{Aa}\eta_{a,l_1}+\lambda_{l_2}^{Aa}\eta_{a,l_2}-q_2^A)\nonumber\\
  &&=(4!)^24(l_1l_2)^2\delta^8(q_1^A+q_2^A).
\end{eqnarray}

\subsection{$D=10$ $\mathcal{N}=1$ SYM}\label{_24}
The $D=10$ $\mathcal{N}=1$ SYM supermultiplet of on-shell states 
consists of the physical polarizations of the gluon $A^{AB'}$ and  the  fermion $\Psi^{A}$ fields.
In this case the on-shell momentum
superspace formalism is not known. The problem is that there are too many $\eta$ variables
\cite{SpinorHelisityForm_D=10Dimentions}
(we need $4$ $\eta$ variables to accommodate all $2^4$ states  in the theory, but the smallest representation of the little group $SO(8)$ gives $8$).

However, one can use the indirect symmetry arguments (see the next section) to show that the ratio of $\mathcal{A}_4^{(L)}/\mathcal{A}_4^{(0)}$ in $D=10$ $\mathcal{N}=1$ SYM has the form similar to that in the $D=4,6,8$ SYM theories. One can also use an alternative formulation of the amplitudes in the $D=10$ $\mathcal{N}=1$ SYM theory based on the pure spinor formalism \cite{PureSpinorsMarfa} to show that at one and two loops the integrand of the ratio of $\mathcal{A}_4^{(L)}/\mathcal{A}_4^{(0)}$ in $D=10$ $\mathcal{N}=1$ SYM indeed has the form identical to that in the  $D=4$ case (also component unitarity based computations are available up to five loops \cite{DixonBDS45loops}).
This strongly supports the above mentioned claim.

\section{The $A_{4}$ amplitude in SYM theories in various dimensions}\label{s1}

\subsection{From physical to colour ordered partial amplitudes}
The aim is to calculate the multiparticle amplitudes on mass shell. For this purpose, we first perform the color decomposition extracting the color ordered partial amplitude~\cite{Reviews_Ampl_General}. The relations between physical and colour ordered  amplitudes look like:
\begin{equation}
\mathcal{A}_n^{a_1\dots a_n,phys.}(p_1^{\lambda_1}\dots p_n^{\lambda_n})=\sum_{\sigma \in S_n/Z_n}Tr[\sigma(T^{a_1}\dots T^{a_n})]
\mathcal{A}_n(\sigma(p_1^{\lambda_1}\dots p_n^{\lambda_n}))+\mathcal{O}(1/N_c).
\end{equation}
The colour ordered amplitude $A_n$ is evaluated in the planar limit which corresponds to $N_c\to \infty$, $g^2_{YM}\to 0$ and $g^2_{YM}N_c$ - fixed.

For the four-point  amplitudes the colour decomposition reduces to
\begin{eqnarray}
\mathcal{A}_4^{a_1\dots a_4,(L),phys.}(1,2,3,4)=T^1\mathcal{A}_4^{(L)}(1,2,3,4)+T^2\mathcal{A}_4^{(L)}(1,2,4,3)+
T^3\mathcal{A}_4^{(L)}(1,4,2,3)
\end{eqnarray}
where $T^i$ denote the trace combinations of $SU(N_c)$ generators in the fundamental representation 
\begin{gather}
T^1=Tr(T^{a_1}T^{a_2}T^{a_3}T^{a_4})+Tr(T^{a_1}T^{a_4}T^{a_3}T^{a_2}),\nonumber \\
T^2=Tr(T^{a_1}T^{a_2}T^{a_4}T^{a_3})+Tr(T^{a_1}T^{a_3}T^{a_4}T^{a_2}),\\
T^3=Tr(T^{a_1}T^{a_4}T^{a_2}T^{a_3})+Tr(T^{a_1}T^{a_3}T^{a_2}T^{a_4}).\nonumber
\end{gather}

What is essential and becomes obvious using the superspace formalism, the four point tree-level amplitude always factorizes
so that the colour decomposed $L$-loop amplitude can be written in the form
\beq
\mathcal{A}_4^{(L)}(1,2,3,4)=\mathcal{A}_4^{(0)}(1,2,3,4)M_4^{(L)}(1,2,3,4)=
\mathcal{A}_4^{(0)}(1,2,3,4)M_4^{(L)}([1+2]^2,[2+3]^2)\nonumber
\eeq
or using the standard Mandelstam variables
\beq
\mathcal{A}_4^{(L)}(1,2,3,4)=\mathcal{A}_4^{(0)}(1,2,3,4)M_4^{(L)}(s,t)
\eeq
It is this $M_4^{(L)}(s,t)$ factorized amplitude that we calculate in this paper.

In general $M_4^{(L)}$ has the form
\beq
M_4^{(L)}(s,t)=(-g^2)^L \sum_{i} \mbox{coef}_i \times \mbox{MasterIntegral}_i,
\eeq
where $g^2 \equiv \frac{g_{YM}^2N_c}{(4\pi)^{D/2}}$, the $\mbox{coef}_i$ are some monomials of $s$ and $t$, the $\mbox{MasterIntegral}_i$ is one of the master integrals in the $D$-dimensional Minkowski space to be evaluated.

\subsection{Dual conformal invariance and the universal expansion} \label{_26}

The $D=4$ $\mathcal{N}=4$ SYM planar S-matrix in addition to the $PSU(2,2|4)$ (super)conformal symmetry
has a new type of symmetry, namely the dual (super)conformal symmetry. One can think of this
symmetry as of (super)conformal transformations  acting on the new dual variables $x^{AB'}_i$ and their fermionic counterparts. The dual variables $x^{AB'}_i$ are defined in $D=4,6,8,10$ dimensions as
\begin{eqnarray}
p_{i}^{AB'}&=&x_i^{AB'}
-x_{i+1}^{AB'}.
\end{eqnarray}
The explicit form of the generators of the dual (super) conformal transformations in $D=4$
as well as details specific for $D=4$ $\mathcal{N}=4$ SYM 
can be found in \cite{DualConfInvForAmplitudesCorch}. 

The dual (super)conformal symmetry is exact at the tree level for a general kinematical configuration.
For the amplitudes at the loop level the dual conformal symmetry is, in general, broken due to the presence of the IR divergences (see the details, for example,  in \cite{BeisertYangianRev}).
However, this symmetry is still \emph{exact} if one considers not the loop amplitudes themselves but rather their \emph{integrands}. For the four-point amplitude this statement mani\-fests itself in the fact that the ratio $\mathcal{A}_4^{(L)}/\mathcal{A}_4^{(0)}$ is given by  the linear
combination of the so-called \emph{dual (pseudo)conformal integrals} with the coefficients given by the rational functions of the Mandelstam variables. 

Remarkable that one can define, at least, the bosonic part of  the dual conformal transformations for the $D=10$ $\mathcal{N}=1$ SYM \cite{SpinorHelisityForm_D=10Dimentions}. Moreover,
it is claimed that the tree level S-matrix of the $D=10$ $\mathcal{N}=1$ SYM is
\emph{covariant} with respect to the dual conformal transformations similar to the
$D=4$ case. This would immediately imply the dual conformal covariance
of the tree level S-matrix for the  $D=4,6,8$ SYM theories since they can be obtained by dimensional reduction \cite{SpinorHelisityForm_D=10Dimentions,ZBern_GenUnit_D=6Helicity,N=4Higgs} from the $D=10$ $\mathcal{N}=1$ SYM. This would also imply that the structure of unitarity cuts in all SYM theories is identical. Combining all these statements together one can conclude  that the integrands of the ratio $\mathcal{A}_4^{(L)}/\mathcal{A}_4^{(0)}$ in the $D=6,8,10$ SYM theories
have the form identical to those of the $D=4$ SYM theory.

Separate investigation of the dual conformal invariance in the $D=6$ $\mathcal{N}=(1,1)$ SYM \cite{D6_DualConformal_Invariance}
as well as explicit computations of the four-point amplitudes up to 3 loops and
the  spinor formalism based results up to 2 loops in $D=10$ \cite{PureSpinorsMarfa} (see also discussion of the $N=1$ $D=10$ unitarity cuts for the 5 loop integrals in \cite{DixonBDS45loops}) strongly support that the integrands
in the $D=4,6,8,10$ dimensional SYM theories are indeed identical.

Consider as an example the four point amplitude in $D=4,6,8,10$ SYM theory. At the one loop the
contribution to $\mathcal{A}_4^{(1)}/\mathcal{A}_4^{(0)}$  is given by the box diagram with the integrand
 \beq
BoxIntegrand=\frac{st}{k^2(k-p_1)^2(k+p_2)^2(k+p_2+p_3)^2}.
 \eeq
Introducing the dual coordinates $x_i$ as (here we omit the spinor indices for simplicity)
$$
p_1=x_{12},~p_2=x_{23},~p_3=x_{34},~p_{4}=x_{41},~k=x_{15},
$$
we can rewrite the integrand of this box diagram as follows:
\beq
BoxIntegrand=\frac{x_{13}^2x_{24}^2}{x_{15}^2 x_{25}^2 x_{35}^2 x_{45}^2}.
\eeq
This expression is invariant under the dual conformal transformations, except inversions and transforms covariantly under the dual conformal inversions $I$ with the weight $(x_5^2)^4$  which is absorbed into the integration measure in $D=4$. In higher dimensions this factor is left behind, however, it does not depend on external  coordinates. 
The requirement of the dual conformal invariance is enough to uniquely fix the combination $x_{13}^2x_{24}^2/x_{15}^2 x_{25}^2 x_{35}^2 x_{45}^2$, i.e. the integrand written in terms of the Box integral $1/x_{15}^2 x_{25}^2 x_{35}^2 x_{45}^2$ with the fixed coefficient $x_{13}^2x_{24}^2$. Similar logic is also true in higher orders of PT.

In combination with the unitarity based method the dual conformal invariance allows one to explicitly express the amplitude
$\mathcal{A}_4^{(L)}/\mathcal{A}_4^{(0)}$ in terms of  some dual (pseudo)conformal integrals with the known coefficients up to $6$ loops.
The all loop expression known as the BDS ansatz  for
$\mathcal{A}_4/\mathcal{A}_4^{(0)}$ was also obtained \cite{BDS}. 

We base our calculation of the four-point amplitude in various dimensions
on  a universal expansion which, as it was mentioned above, is valid  for any D. The difference is only the dimension of the integration while the integrands stay universal.

The expansion  for  the ratio $\mathcal{A}_4/\mathcal{A}_4^{(0)}$ up to 3 loops is schematically presented below
\beq
\begin{tabular}{c}
$\dfrac{\mathcal{A}_4}{\mathcal{A}_4^{(0)}}=1+\sum\limits_L M^{(L)}_4(s,t)=$  \\
$\includegraphics[scale=0.35]{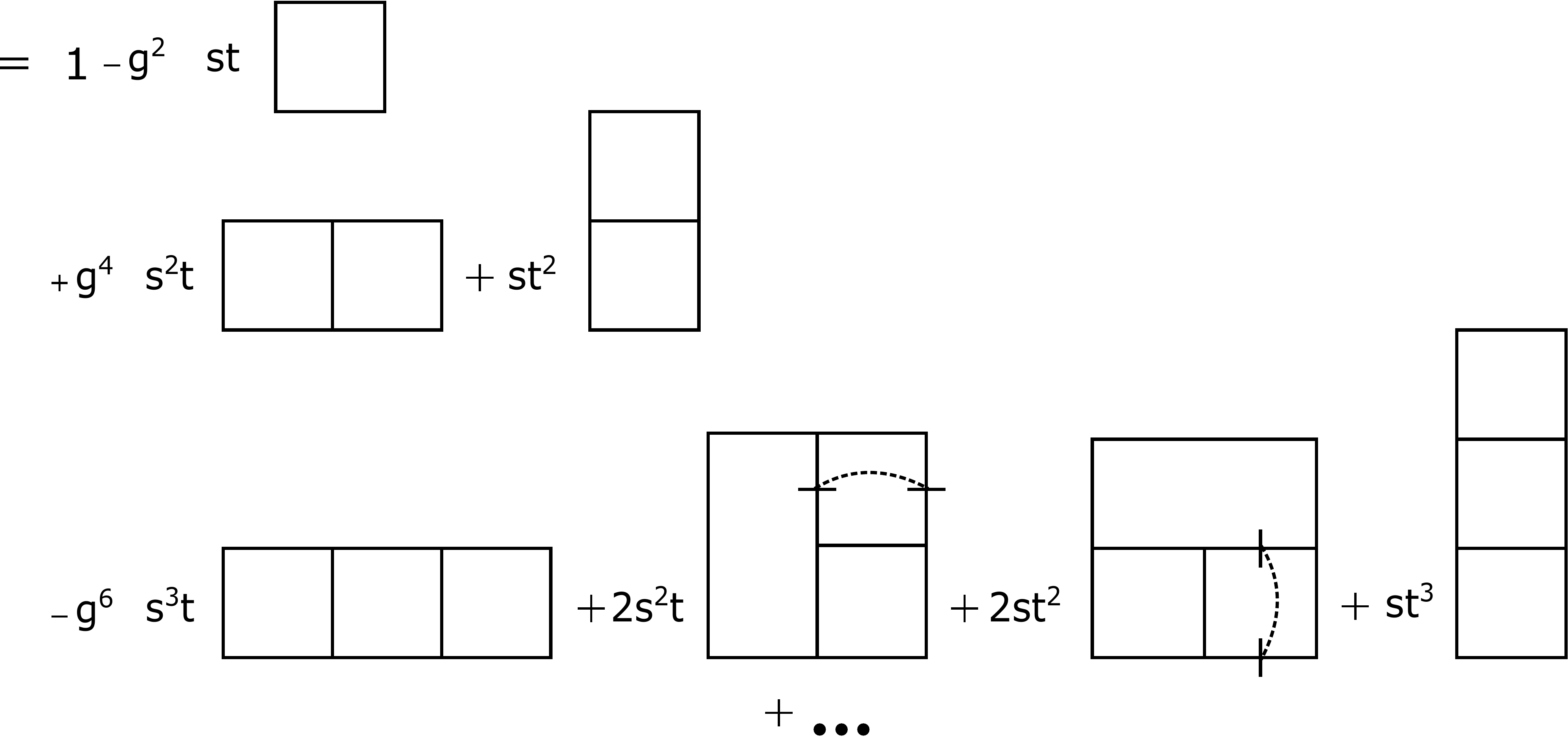}$ \end{tabular} \label{expan}
\eeq
where the connected strokes on the lines mean  the square of the flowing momentum.
In what follows, we consider the $D$=6, 8 and 10 cases.

\section{Calculation of Integrals}
\subsection{Analytical evaluation}
Due to eq.(\ref{expan}) the problem of calculation of divergences is reduced to the scalar master integrals, which are universal for any dimension. To evaluate them, we use the dimensional regulari\-zation. 
Throughout the paper we accept the following definition of the $L$-loop master integrals:
\begin{eqnarray}\label{2}
\mbox{MasterIntegral}_i=\left(\frac{1}{i\pi^{D/2}}\right)^L\int d^Dk_1...d^Dk_L \frac{Num_i.}{Den_i.}.
\end{eqnarray}

Since we are interested only in the leading divergences (the leading poles in dimensional regularization), the task is essentially simplified. One has to admit that to calculate the leading pole, there is no need to calculate the multiloop diagram itself. The leading pole follows from the lowest order singularity due to the nature of the ${\cal R}$-operation. It is valid in any local field theory and guarantees the locality of divergences if the lower order counterterms are taken into account. 

Let us briefly recall the main notions of the $\R$-operation~\cite{Bogolubov,Zavialov}. Being applied to any Green function  $\Gamma$ (or any particular graph $G$, as in our case) it subtracts all the UV divergences including those of divergent subgraphs and leaves the finite expression. The use of the $\R$-operation is equivalent to addition of the counter terms to the initial Lagrangian. The $\R$ operation can be written in terms of the subtraction operators in the factorized form
\beq  \R G = \prod_\gamma (1-K_\gamma)G, \eeq
where the subtraction operator $K_\gamma$ subtracts the UV divergence of a given subgraph $\gamma$ (for the minimal subtraction scheme  the operator $K$ singles out the  $1/\epsilon^n$ terms) and the product goes over all divergent subgraphs including the graph itself.

It is useful to define also the incomplete $\R$ operation denoted by $\R'$ which subtracts only  the subdivergences of the graph $G$. The full $\R$ operation is then defined as
\begin{equation}
\R G = (1-K) \R' G,
\end{equation}
where $K$ without subscript is $K_G$. The $K\R' G$  is the counter term corresponding to the  graph $G$.
Each counter term contains only the superficial divergence and is  {\it local} in coordinate space (in our case it must be a polynomial of external momenta). 

The $\R'$ operation for any graph $G$ can be defined by the forest formula, but for our calculations it is more convenient to use  the recursive definition via the $\R'$ operation for  divergent subgraphs (for details and examples see chapter 3 in \cite{Vasiliev}):
\begin{equation}
{\cal R}' G= \left(1-\sum_\gamma (K{\cal R}')_\gamma +\sum_{\gamma,\gamma'}(K{\cal R}')_\gamma \; (K{\cal R}'_{\gamma'}) - ...\right) G,
\label{str}
\end{equation}
The sum goes over all 1PI, UV-divergent subgraphs of the given diagram and the multiple sums include only the non-intersecting subgraphs. 
And 
\begin{equation}
(K\R')_\gamma \;G = K\R'(\gamma) * G/\gamma,
\end{equation}
where $G/\gamma$ denotes the graph $G$ with the subgraph $\gamma$ shrinked to point, the $*$ operation inserts the subgraph's counter term  (in our case it is a  polynomial of momenta external to a given subgraph)  into the remaining graph $G/\gamma$.


When applying this formula to the diagrams at hand one finds out that for the n-loop diagram the ${\cal R}'$-operation results in the series of terms (we consider only the leading pole)
\beq
\frac{A_n (\mu^2)^{n\epsilon}}{\epsilon^n}+\frac{A_{n-1}(\mu^2)^{(n-1)\epsilon}}{\epsilon^n}+ ... +\frac{A_1(\mu^2)^{\epsilon}}{\epsilon^n},\label{Rn}
\eeq
where the term like $\frac{A_{k}(\mu^2)^{k\epsilon}}{\epsilon^n}$ comes from the $k$-loop graph which survives after subtraction of the $(n-k)$-loop counterterm. 
 The full expression (\ref{Rn}) has to be local, i.e. should not contain terms like $(\log \mu)^k/\epsilon^m$ for all $k,m>0$ while being expanded over $\epsilon$. (For simplicity hereafter we put $\mu^2 \equiv \mu$.)
This requirement gives us $n-1$ equations for the coefficients $A_i$. Solving them one gets
\beq
A_n=(-1)^{n+1}\frac{A_1}{n}. \label{rel}
\eeq
In the case when the first divergence appears at $k$ loops (as in the D=6 case where k=3) this formula is slightly modified and looks like
\beq
A_n=(-1)^{n+k}\frac{(k-1)!(n-k-1)!}{n(n-2)!}A_1. \label{rel2}
\eeq
This means that performing the  ${\cal R}'$-operation one can take care only of the one loop  diagrams
surviving after contraction and get the desired pole term via eq.(\ref{rel2}). This observation drastically simplifies the calculation of the leading pole.

\begin{figure}[!h]
  \centering
  \includegraphics[scale=0.5]{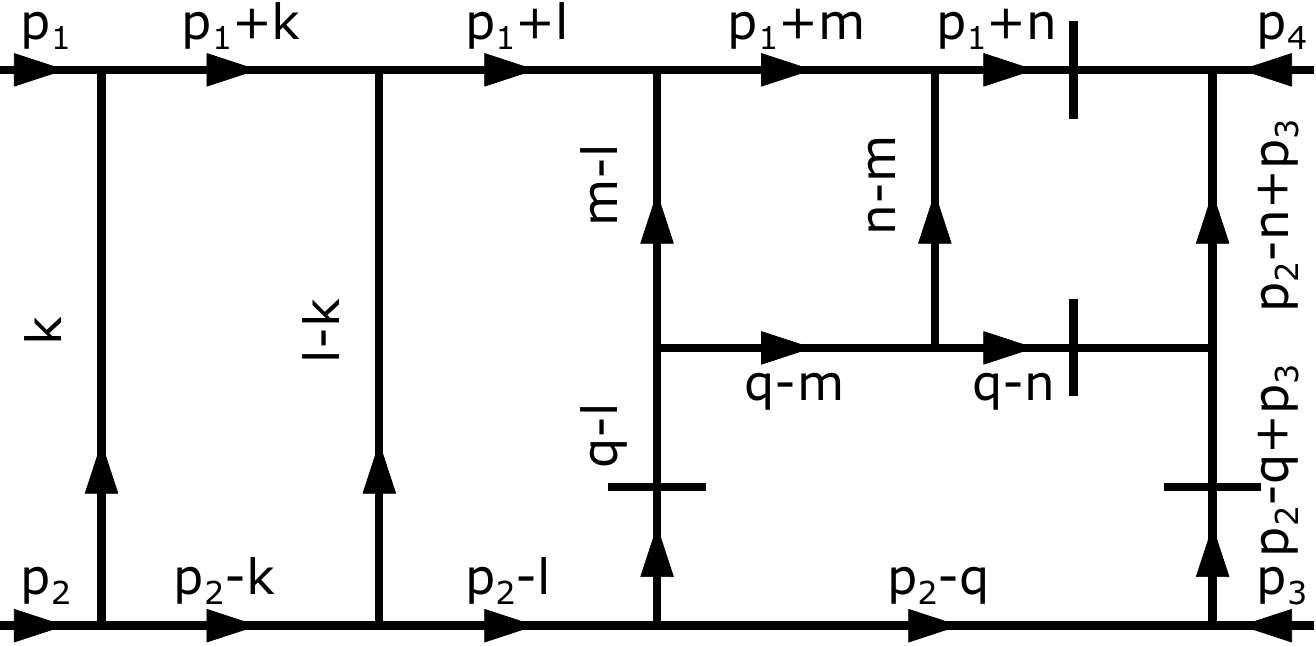}
  \caption{The choice of the inner momenta}\label{momen}
\end{figure}

To demonstrate how this technique works, we calculate the leading pole of the  5-loop diagram $I_4^{(5)}$ in D=6 (see Appendix A). For pedagogical purposes we describe first the calculation using the full ${\cal R}'$-operation and then show how the truncated version using eq.(\ref{rel2}) works. At the beginning, we define the inner momenta, as shown in Fig.\ref{momen}.
The slashes on the lines correspond to the numerator
\beq
Num = (p_2-l+p_3)^2 (p_1+q)^2 \label{num}
\eeq
It is useful to rewrite the first bracket as $(p_2-l+p_3)^2 = (p_2-l)^2 + 2p_3(p_2-l)+p_3^2$. Then, having in mind that $p_3^2=0$  we have two terms. 

\begin{figure}[!h]
  \centering
  \includegraphics[scale=0.4]{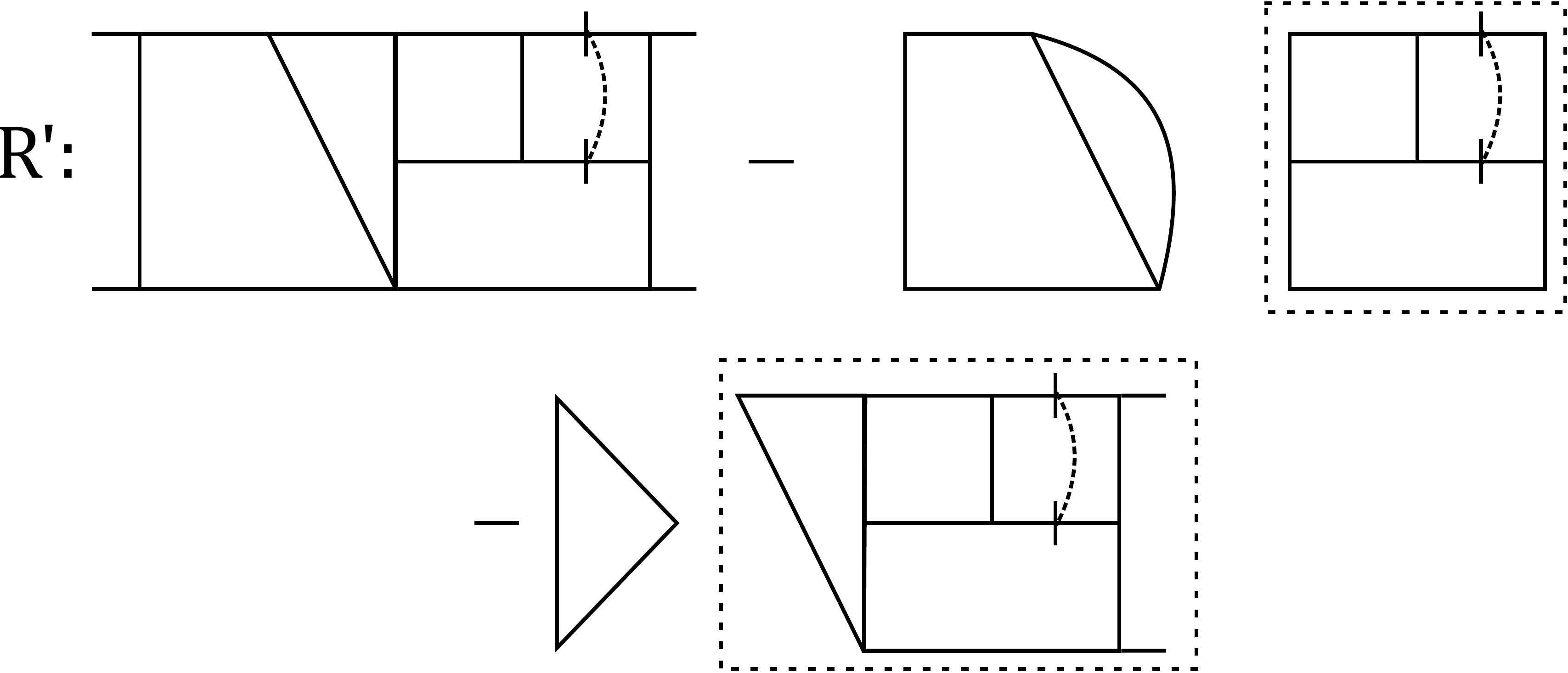}
  \caption{The ${\cal R}'$-operation for the first term}\label{R_operation5_2}
\end{figure}

In the first term the numerator $(p_2-l)^2$  cancels one of the propagators. The ${\cal R}'$-operation for this diagram is shown in Fig.\ref{R_operation5_2}. There are two 1PI divergent subgraphs, the three-loop and the four-loop ones. Since one subgraph is inside the other the ${\cal R}'$-operation contains only the first sum in eq.(\ref{str}). For the graphs framed in dashed boxes one has to take the ${\cal KR}'$ expression.

We start from the 3-loop subgraph. Since the singular part of the tennis-court 3-loop graph  equals  $-1/6\epsilon$ and does not depend on momenta, one actually has to calculate the remaining
2-loop graph. To do this, we notice that the one-loop bubble in D=6 is proportional to the ingoing momentum squared, so when substituting it into the 2-loop graph one propagator will be canceled and the resulting graph will take the form of a bubble with ingoing momentum equal to $p_2$. Since this bubble is also proportional to the square of the ingoing momentum and $p_2^2=0$, this leads the contribution equal to zero.

We now repeat the same procedure for the 4-loop subgraph. To calculate the ${\cal K R}'$-operation for it, we have again to substitute the 3-loop tennis-court graph, shrink it to a point to get the bubble, which again  is proportional to the incoming momentum squared. This momentum is not on shell but being substituted into the remaining triangle in Fig.\ref{R_operation5_2} cancels one propagator. This converts the triangle into the bubble which for the same reason as above is equal to zero.

Thus, our conclusion is that the first term gives zero contribution.

\begin{figure}[!h]
  \centering
  \includegraphics[scale=0.4]{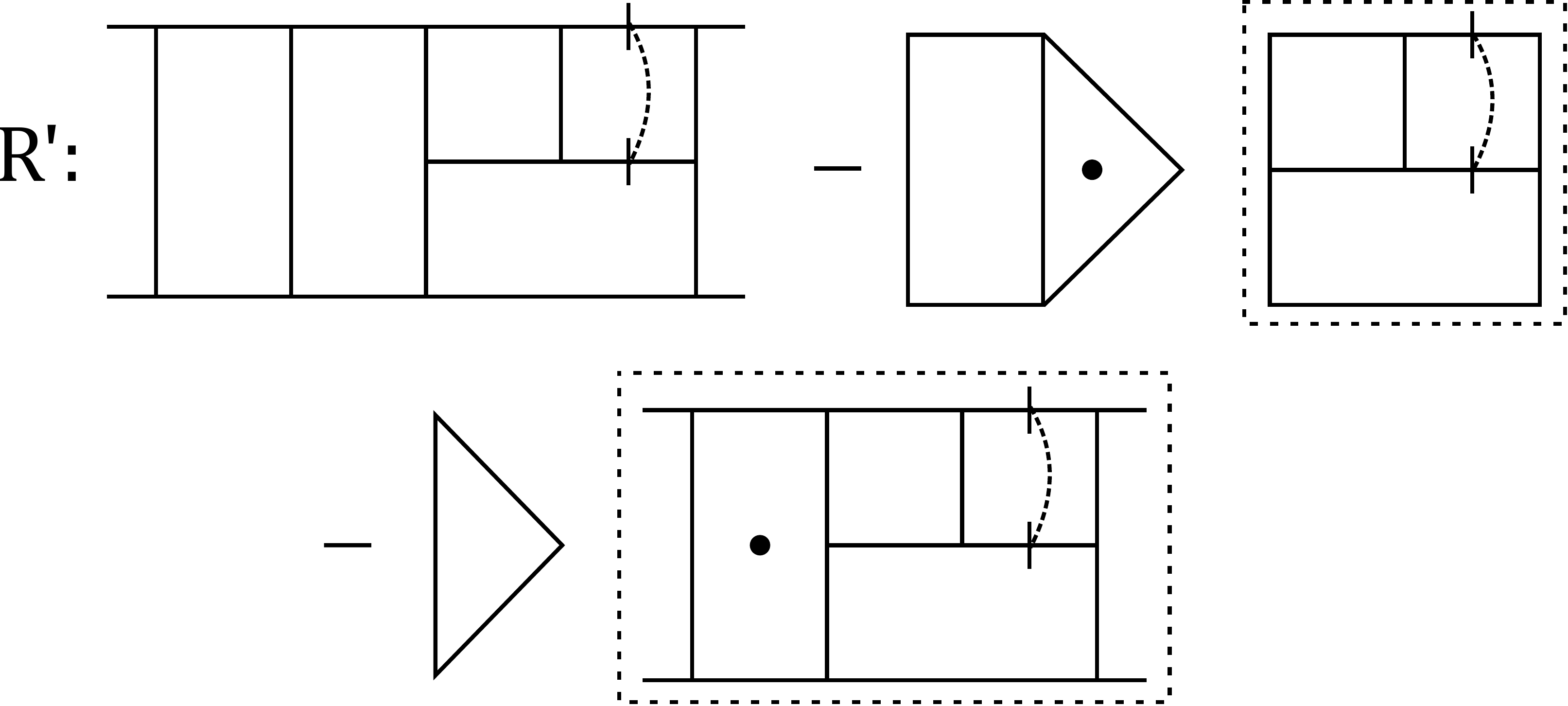} 
  \caption{The ${\cal R'}$-operation for the second term. The dot corresponds to the numerator $2p_3(p_2-l)$}\label{R_operation5_1}
\end{figure}

\begin{figure}[!h]
  \centering
  \includegraphics[scale=0.4]{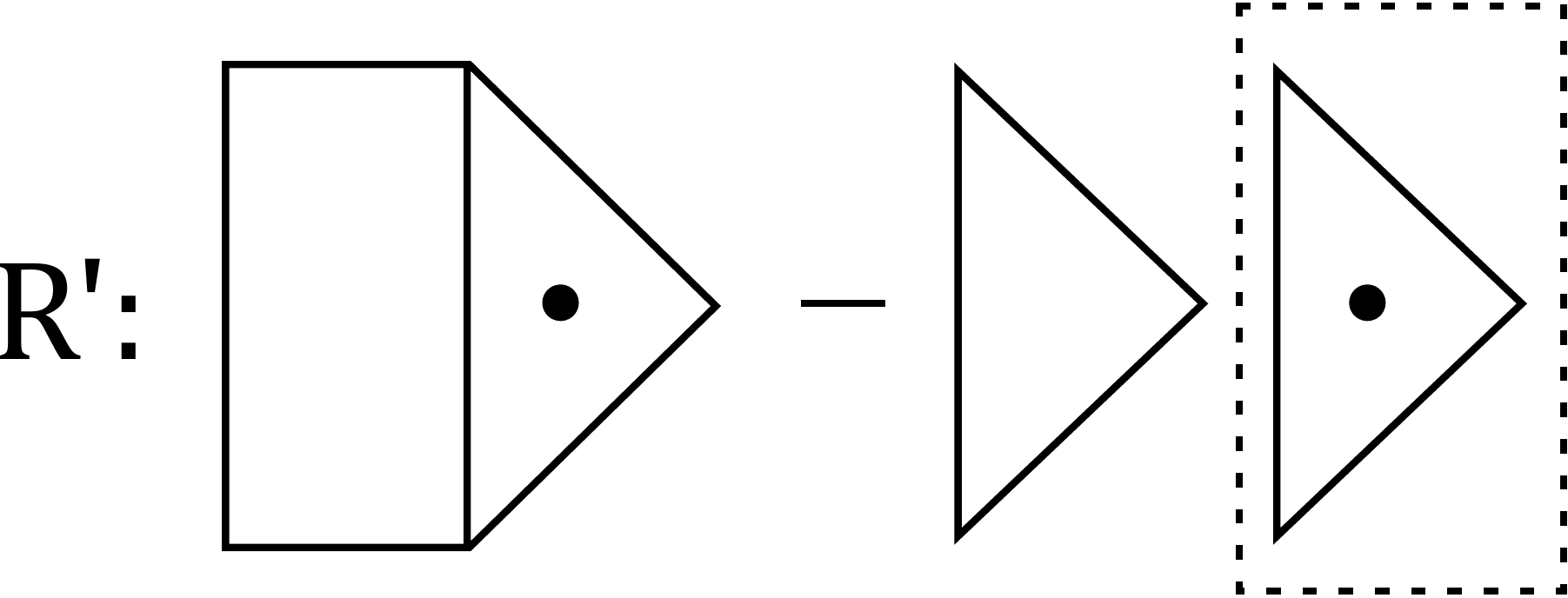}
  \caption{The ${\cal R'}$-operation for the 2-loop subgraph}\label{R_operation2}
\end{figure}

Consider now the second term. The ${\cal R}'$-operation in this case is shown in Fig.\ref{R_operation5_1}. 
Again we have two divergent subgraphs, one inside the other, which give two contributions. The first one contains the 2-loop graph. Its singular part can be calculated via the ${\cal R}'$-operation shown in Fig.\ref{R_operation2}.
The divergent subgraph here is a triangle
\beq
\int \frac{2p_3(p_2-l)\ d^{6-2\epsilon}l}{(p_2-l)^2(l-k)^2(p_1+l)^2} = -\frac{1}{6\epsilon}2p_3(2p_2-k+p_1) + finite\ terms \label{tri}
\eeq
Substituting it into the remaining triangle obtained by shrinking the first one to a point we get the integral
\beq
\int \frac{2p_3(2p_2-k+p_1)\ d^{6-2\epsilon}k}{(p_2-k)^2(k)^2(p_1+k)^2} =-\frac{1}{6\epsilon}2p_3(5p_2+4p_1) + finite\ terms \label{tri2}
\eeq
Now we have for the 2-loop graph
\beq
{\cal R'}: \frac{A_2\mu^{2\epsilon}}{\epsilon^2}-\frac{2p_3(5p_2+4p_1)\mu^\epsilon}{36\epsilon^2}.
\eeq
From this equation requiring the cancellation of the $\log\mu/\epsilon$ term we get
\beq
A_2=\frac{5t+4u}{72\epsilon^2}=\frac{t-4s}{72\epsilon^2}.
\eeq
Thus, the first contribution to the ${\cal R}'$-operation is
\beq \left(-\frac{1}{6\epsilon}\right)\frac{(t-4s)\mu^{2\epsilon}}{72\epsilon^2}=-\frac{(t-4s)\mu^{2\epsilon}}{216\cdot 2 \epsilon^3}
\label{res1}.
\eeq

\begin{figure}[!h]
  \centering
  \includegraphics[scale=0.4]{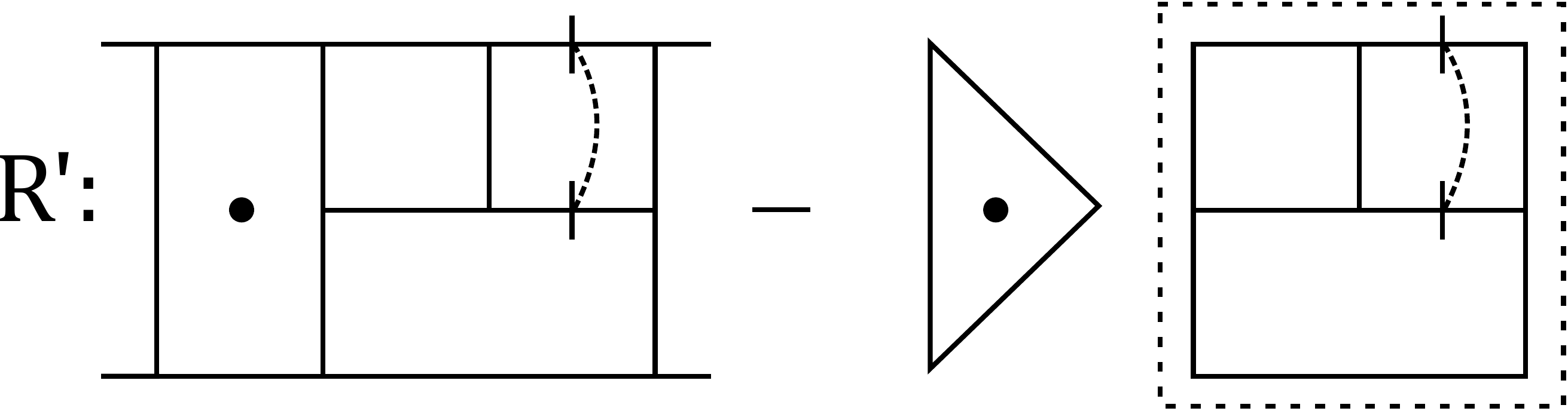}
  \caption{The ${\cal R'}$-operation for the 4-loop subgraph}\label{R_operation4_1}
\end{figure}

For the second contribution
the problem is reduced to the 4-loop counterterm. To compute it we again use the the ${\cal R}'$-operation.  The corresponding graphs are shown in Fig.\ref{R_operation4_1}.
The calculation repeats the one performed above (\ref{tri}) and one gets  the ${\cal R'}$-operation for the 4-loop subgraph 
\begin{gather}
{\cal R'}: A_4\frac{\mu^{4\epsilon}}{\epsilon^2}-\left(-\frac{1}{6\epsilon}\right)\left(-\frac{\mu^{\epsilon}}{6\epsilon}2p_3(2p_2-k+p_1)\right).
\end{gather}
This gives for $A_4$
\beq
A_4 = \frac{2p_3(2p_2-k+p_1)}{4\cdot36}. 
\eeq
Thus, the ${\cal K R'}$ for the 4-loop subgraph is
\beq
{\cal K R'} = \frac{2p_3(2p_2-k+p_1)}{4\cdot 36\epsilon^2}\mu^{4\epsilon}- 
\frac{2p_3(2p_2-k+p_1)}{36\epsilon^2}\mu^\epsilon=-3 \frac{2p_3(2p_2-k+p_1)}{4\cdot 36\epsilon^2}
\eeq
Next, we substitute the obtained coefficient into the  second part of the ${\cal R'}$-operation  for the  5-loop subgraph Fig.\ref{R_operation2}. The resulting triangle integral including the numerator has the form (\ref{tri2}) which gives 
\beq
\left(-\frac{3\mu^\epsilon}{4\cdot 36\epsilon^2}\right)\left(-\frac{1}{6\epsilon}2p_3(5p_2+4p_1)\right)=\frac{\mu^\epsilon 3(t-4s)}{4\cdot 216 \epsilon^3}.\label{res2}
\eeq
Summing up eqs.(\ref{res1},\ref{res2}) for the full ${\cal R'}$-operation for the 5-loop graph one has
\beq
{\cal R'}: \frac{A_5\mu^{5\epsilon}}{\epsilon^3}+\frac{\mu^{2\epsilon}(t-4s)}{2\cdot 216 \epsilon^3}
-\frac{\mu^\epsilon 3(t-4s)}{4\cdot 216 \epsilon^3}
\eeq
expanding over $\epsilon$ and collecting the terms of $\log\mu/\epsilon^2$ and $\log^2\mu/\epsilon$ we get two equations to determine $A_5$:
\beqa
\log\mu: & 5A_5+2/2/216(t-4s)-3(t-4s)/4/216=0 \\
\log^2\mu: & 25A_5+4/2/216(t-4s)-3(t-4s)/4/216=0
\eeqa
Solution to these eqs is
\beq
A_5=-\frac{(t-4s)}{20\cdot 216}=\frac{s-t/4}{30\cdot 36}
\eeq
Consistency of the two equations serves as a check of correctness of the calculations.

One can arrive at the same result in a shorter way using the truncated ${\cal R'}$-operation and evaluating only the term with the one-loop subgraph (\ref{res2}). Indeed, using relation (\ref{rel2}) for $n=5, k=3$ one gets
\beq
A_5=\frac{1}{15}A_1=-\frac{1}{15}\frac{3(t-4s)}{4\cdot 216}=
\frac{s-t/4}{30\cdot 36}.\label{fulres}
\eeq
This gives us the result listed in  Table \ref{Answers1_4}. 

\begin{figure}[!h]
  \centering
  \includegraphics[scale=0.5]{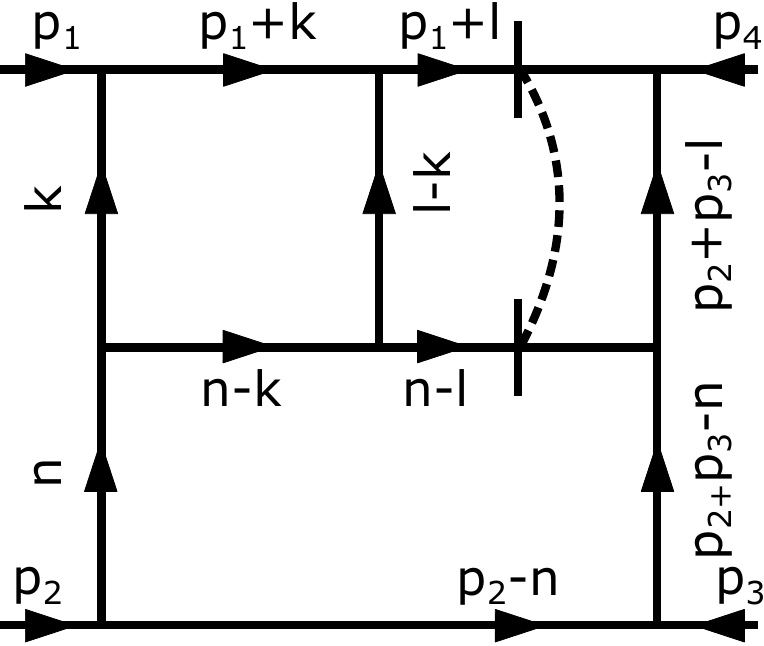}
  \caption{The choice of the inner momenta}\label{momen3}
\end{figure}

As the second example we consider the evaluation  of the the diagram $I_2^{(3)}$ in D=8 by means of the truncated ${\cal R'}$-operation.   Again,  we define the inner momenta, as shown in Fig.\ref{momen3}.
Keeping in mind that $p_i^2=0$ we rewrite the numerator in the following way:
\beq \label{num2}
Num=(p_1+n)^2=2(p_1n)+n^2.
\eeq
The term proportional to $n^2$ cancels the  corresponding propagator. Thereby, after shrinking the 2-loop subgraphs we  get the bubble diagram which is proportional to the on-shell momentum squared and is equal to zero.  Thus, the numerator finally gets the form
$ Num = 2(p_1n)$.

\begin{figure}[!h]
  \centering
  \includegraphics[scale=0.5]{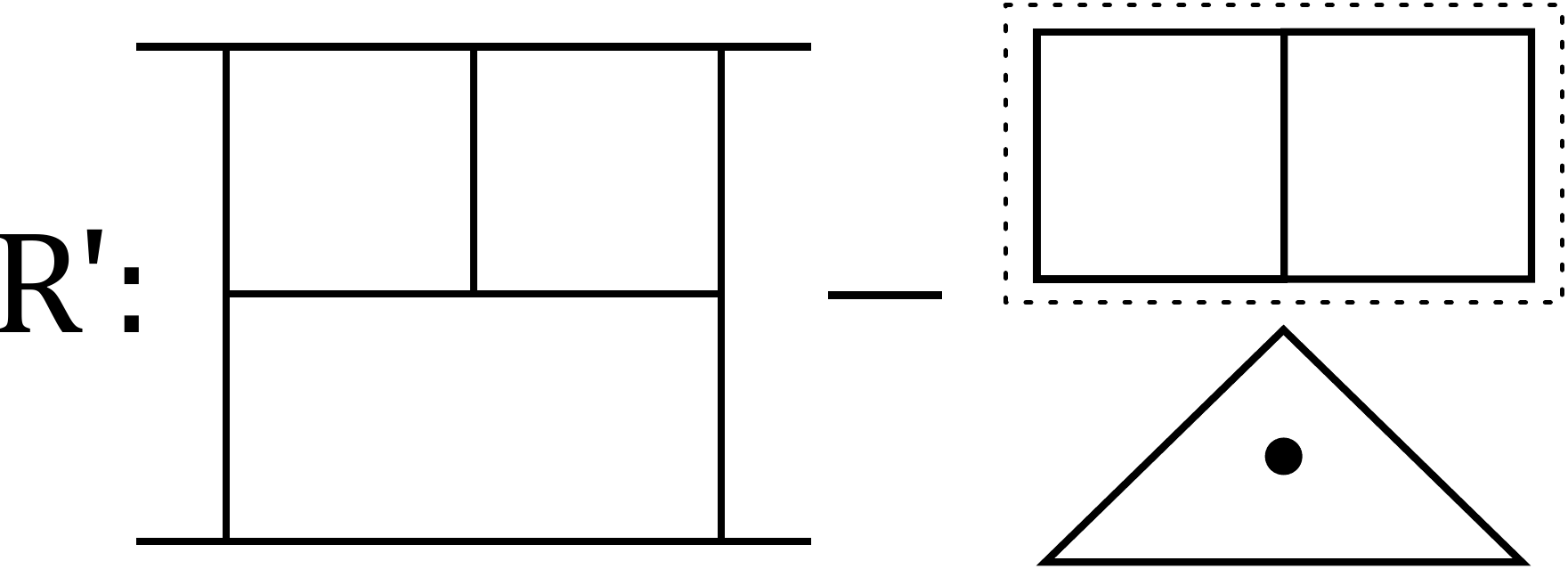}
  \caption{The ${\cal R'}$-operation for the 3-loop graph. The dot corresponds to the numerator  $  2(p_1n)$}\label{R_operation33}
\end{figure}

Consider now the ${\cal R'}$ - operation for this diagram having in mind that we are interested only in the one loop remaining graphs shrinking the rest of the diagram to a point. Since the upper right and left boxes transform into bubbles after shrinking the rest part of the diagram and these bubbles are again  proportional to the on-shell momentum squared, the answer for both of them is zero. Therefore, the only one loop graph that survives is the lower box. The ${\cal R'}$ - operation is shown in Fig.\ref{R_operation33}

\begin{figure}[!h]
  \centering
  \includegraphics[scale=0.5]{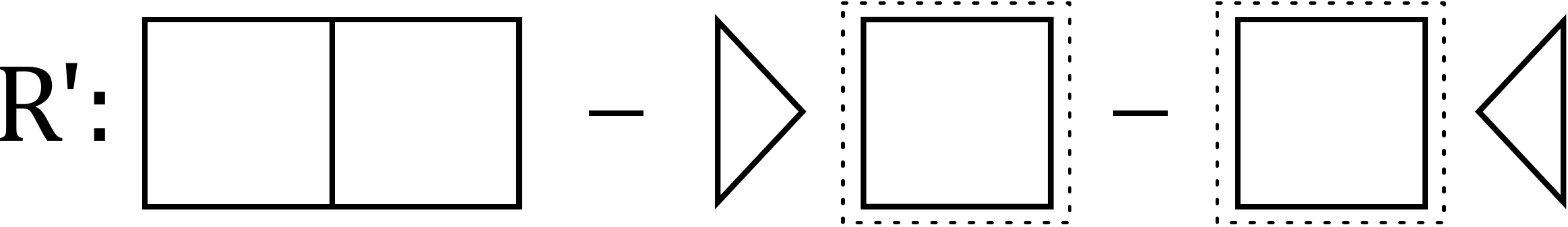}
  \caption{The ${\cal R'}$-operation for the double box}\label{R_operation32}
\end{figure}

The calculation  of the double box subgraph is performed again using the ${\cal R'}$-operation shown in Fig.\ref{R_operation32}
The integral for the left box has the following form: 
\beq
\int \frac{ d^{8-2\epsilon}k}{k^2(n-k)^2(l-k)^2(p_1+k)^2} = \frac{1}{6\epsilon} + finite\ terms \label{box}.
\eeq
Since its singular part does not depend on momenta, the right box  is the same. 
For the first and second triangles in Fig.\ref{R_operation32} the integrals are
\begin{gather}
\int \frac{ d^{8-2\epsilon}k}{k^2(n-k)^2(p_1+k)^2} = -\frac{2n^2+2(p_1n)}{24\epsilon} + finite\ terms, \nonumber \\
\int \frac{ d^{8-2\epsilon}l}{(p_1+l)^2(n-l)^2(p_2+p_3-l)^2} = -\frac{n^2+2\cdot2(p_1n)+2(p_4n)+t}{24\epsilon} + finite\ terms.\label{triangl}
\end{gather}
Substituting expressions (\ref{box},\ref{triangl}) into the  ${\cal R'}$-operation we have for the 2-loops subgraph 
\beq
{\cal R'}: A_2\frac{\mu^{2\epsilon}}{\epsilon^2}-\mu^{\epsilon}\left(-\frac{2n^2+2\cdot2(p_1n)+2(p_4n)+t}{24\epsilon}\right)\frac{1}{6\epsilon}-\mu^{\epsilon}\left(-\frac{2n^2+2(p_1n)}{24\epsilon}\right)\frac{1}{6\epsilon}
\eeq
Evaluating the  coefficient $A_2$ from the requirement of cancelation of the $\log\mu/\epsilon$ term one gets
\beq \label{a2}
A_2 = -\frac{4n^2+3\cdot2(p_1n)+2(p_4n)+t}{2\cdot6\cdot24}.
\eeq
This gives the ${\cal KR'}$ for the double box
\beqa
{\cal KR'} \mbox{Double Box}&= & A_2\frac{\mu^{2\epsilon}}{\epsilon^2}-\left(-\frac{4n^2+3\cdot2(p_1n)+2(p_4n)+t}{2\cdot6\cdot24}\right)\frac{\mu^{\epsilon}}{\epsilon^2} \nonumber \\ &=&\frac{4n^2+3\cdot2(p_1n)+2(p_4n)+t}{2\cdot6\cdot24\ \epsilon^2}\label{DBKR}
\eeqa
We now turn back to the ${\cal R'}$-operation (Fig.\ref{R_operation33}). Substituting  (\ref{DBKR}) and having in mind the numerator $2(p_1n)$ we finally get the triangle with 3 powers of internal momentum in the numerator. Performing this integration one has  
\beq
\frac{A_1}{\epsilon^3}=\frac{t\left(3t^2-2st+s^2\right)}{3!4!5!3\epsilon^3}
\eeq
According to eq.(\ref{rel}), this gives
\beq
A_3=\frac{1}{3}A_1=\frac{t\left(3t^2-2st+s^2\right)}{3!4!5!9}
\eeq
This way we get the expression for the leading pole of the diagram $I_2^{(3)}$.

Applying the described truncated ${\cal R'}$-operation we calculated all the leading poles for the diagrams in $D=6$ up to 5-loops and  in $D=8,10$  up to 4-loops. The results are presented in Tables  \ref{Answers1_4} and \ref{Answers5}.
\begin{table}[!h]
\centering
\begin{tabular}{|c|c|c|c|c|}\hline
MI & Comb & $D=6$ & $D=8$ & $D=10$ \\ \hline \hline
\multirow{2}{*}{ $I_1^{(1)}$} &\multirow{2}{*}{ $st$}&\multirow{2}{*}{conv} & \multirow{2}{*}{$\frac{1}{3!\epsilon}$ }&  \multirow{2}{*}{$\frac{s+t}{5!\eps}$} \\
&&&&\\ \hline \hline
 \multirow{2}{*}{$I_1^{(2)}$}&\multirow{2}{*}{ $s^2t$}&\multirow{2}{*}{conv} & \multirow{2}{*}{$-\frac{s}{3!4!\epsilon^2}$ }& \multirow{2}{*}{$\frac{-s^2(8s+2t)}{5!7!\eps^2}$} \\
&&&&\\ \hline \hline
\multirow{2}{*}{ $I_1^{(3)}$}&\multirow{2}{*}{ $s^3t$}&\multirow{2}{*}{conv} & \multirow{2}{*}{$\frac{s^2}{4!5!\epsilon^3}$ }& \multirow{2}{*}{$\frac{-2s^4(135s+11t)}{5!7!7!3\eps^3}$}\\
&&&& \\ \cline{1-5}
 \multirow{2}{*}{$I_2^{(3)}$}&\multirow{2}{*}{ $2s^2t$}&\multirow{2}{*}{$-\frac{1}{6\epsilon}$  }& \multirow{2}{*}{$\frac{s(3s^2-2st+t^2)}{3!4!5!9\epsilon^3}$} & \multirow{2}{*}{$\frac{-s^2\left(14s^4-10s^3t +\frac{33}{5}s^2t^2 -\frac{19}{5}st^3+\frac{8}{5}t^4\right)}{5!7!7!9\eps^3}$} \\ 
 &&&& \\ \hline \hline
\multirow{2}{*}{ $I_1^{(4)}$}  &\multirow{2}{*}{ $s^4t$}&\multirow{2}{*}{conv} & \multirow{2}{*}{$-\frac{210s^3}{3!4!5!6!\epsilon^4}$} &\multirow{2}{*}{$\frac{-32s^6(99s+2t)}{5!7!7!7!3\eps^4}$}  \\
 &&&& \\  \cline{1-5}
 \multirow{3}{*}{$I_2^{(4)}$}  & \multirow{3}{*}{ $2s^3t$}&\multirow{3}{*}{$\frac{1}{48\epsilon^2}$}&\multirow{3}{*}{ $\frac{s^2\left(-\frac{430}{21}s^2+\frac{4}{9}st-\frac{1}{18}t^2\right)}{3!4!5!6!\epsilon^4}$ } & \multirow{3}{*}{$\frac{-2s^4\left(\begin{smallmatrix}\frac{1502144}{33}s^4-\frac{1085791}{33}s^3t \\ +\frac{2044}{5}s^2t^2 -\frac{1001}{15}st^3+\frac{112}{15}t^4\end{smallmatrix}\right)}{5!7!7!7!7!\eps^4}$} \\
 &&&&  \\
 &&&& \\  \cline{1-5}
 \multirow{3}{*}{$I_3^{(4)}$} &\multirow{3}{*}{ $s^3t$}&\multirow{3}{*}{$\frac{1}{24\epsilon^2} $} &\multirow{3}{*}{ $\frac{s^2\left(-\frac{20}{3}s^2+\frac{8}{9}st-\frac{1}{9}t^2\right)}{3!4!5!6!\epsilon^4}$ }& \multirow{3}{*}{$\frac{-28s^4\left(\begin{smallmatrix}8512s^4-1043s^3t+\frac{876}{5}s^2t^2 - \\ -\frac{143}{5}st^3+\frac{16}{5}t^4\end{smallmatrix}\right)}{5!7!7!7!7!3\eps^4}$} \\ 
  &&&&  \\
 &&&& \\  \cline{1-5}
 \multirow{3}{*}{$I_4^{(4)}$}  &\multirow{3}{*}{ $2s^2t$}&\multirow{3}{*}{$\sim \frac{1}{\epsilon}$ }&\multirow{3}{*}{ $\frac{s\left(\begin{smallmatrix}-\frac{45}{14}s^4+\frac{18}{7}s^3t-\frac{27}{14}s^2t^2 \\ +\frac{9}{7}st^3-\frac{9}{14}t^4\end{smallmatrix}\right)}{3!4!5!6!\epsilon^4}$ }& \multirow{3}{*}{$\frac{-s^2\left(\begin{smallmatrix}-\frac{7504}{1287}s^7+\frac{7819}{1716}s^6t-\frac{1475}{429}s^5t^2+\frac{12745}{5148}s^4t^3  \\ -\frac{716}{429}s^3t^4+\frac{1747}{1716}s^2t^5 - \frac{673}{1287}st^6 +\frac{105}{572}t^7 \end{smallmatrix}\right)}{5!7!7!7!\eps^4}$}  \\
  &&&&  \\
 &&&& \\  \cline{1-5}
 \multirow{3}{*}{$I_5^{(4)}$}  &\multirow{3}{*}{ $4s^2t$}&\multirow{3}{*}{$\frac{t-s}{3\cdot48\epsilon^2}$  }&\multirow{3}{*}{ $\frac{s\left(\begin{smallmatrix}-\frac{15}{28}s^4+\frac{25}{63}s^3t-\frac{65}{252}s^2t^2 \\ + \frac{5}{42}st^3-\frac{1}{28}t^4\end{smallmatrix}\right)}{3!4!5!6!\epsilon^4}$ }&\multirow{3}{*}{$\frac{-4s^2\left(\begin{smallmatrix}-\frac{95200}{143}s^7+\frac{67634}{143}s^6t-\frac{225008}{715}s^5t^2+\frac{136514}{715}s^4t^3  \\ -\frac{6608}{65}s^3t^4+\frac{6706}{143}s^2t^5 - \frac{7420}{429}st^6 +\frac{1715}{429}t^7 \end{smallmatrix}\right)}{5!7!7!7!7!\eps^4}$}  \\
 &&&&  \\
 &&&& \\  \cline{1-5}
\end{tabular}\vspace{0.2cm}
\caption{The leading poles of the diagrams  up to 4-loops  for $D = 6,8$ and $10$} \label{Answers1_4}
\end{table}
\begin{table}[!h]\vspace{0.9cm}
\centering
\begin{tabular}{|c|c|c|c|c|}\hline
MI &$I_1^{(5)}$  &  $I_2^{(5)}$ & $I_3^{(5)}$ & $I_4^{(5)}$ \\ \cline{1-5}
Comb &$2s^4t$ &$2s^4t$ &$4s^3t$ &$2s^3t$\\ \cline{1-5}
&&&& \\
\multirow{-2}{*}{Int}& \multirow{-2}{*}{$-\frac{1}{\epsilon^3}\frac{3}{36\cdot40}$} & \multirow{-2}{*}{$-\frac{1}{\epsilon^3}\frac{9}{36\cdot40} $ } & \multirow{-2}{*}{$\frac{1}{\epsilon^3}\frac{s-t/4}{36\cdot15}$ }  & \multirow{-2}{*}{$\frac{1}{\epsilon^3}\frac{s-t/4}{36\cdot30}$ } \\ \hline  \hline
MI &$I_5^{(5)}$ &$I_6^{(5)}$ &$I_7^{(5)}$ &$I_8^{(5)}$ \\ \cline{1-5}
comb &$4s^2t$ &$2s^2t$ &$4s^2t$ &$4s^2t$\\ \hline
&&&&\\
\multirow{-2}{*}{Int} & \multirow{-2}{*}{$-\frac{1}{\epsilon^3}\frac{s^2-st+t^2}{36\cdot80} $}& \multirow{-2}{*}{$-\frac{1}{\epsilon^3}\frac{s^2-st+t^2}{36\cdot40} $}  & \multirow{-2}{*}{$\frac{1}{\epsilon^3}\frac{s^2-st+t^2/3}{36\cdot80} $} &\multirow{-2}{*}{$\frac{1}{\epsilon^3}\frac{s^2-st+t^2/3}{36\cdot80}$} \\ \cline{1-5}
 \end{tabular}\vspace{0.2cm}
 \caption{The  leading poles of the diagrams in 5-loops  for $D=6$}
 \label{Answers5}
\end{table}

\subsection{Numerical evaluation}
Since in higher loops the evaluation even of the leading pole happens to be a complicated task, we performed a numerical check of our calculations. To evaluate the graph contributi\-ons to the leading pole numerically, we first go to the Euclidean space and then  use the alpha-representation and the method of sector decomposition \cite{heinrich}.

For  the diagrams under consideration  the leading poles ($L.P.$)  are polynomials of $s$ and $t$. For the diagram $G$ with the degree of divergence equal to $2N$ the leading pole is 
\begin{equation}
L.P.(G(s,t))= \sum\limits_{i=0}^N C_{i,N-i} \; s^i \; t^{N-i}
\end{equation}
The coefficients $C_{i,j}$ can be calculated performing the differentiation of the integrand over $s$ and $t$ 
\begin{equation}
L.P.(G(s,t))= \sum\limits_{i=0}^N \; s^i \; t^{N-i} L.P.( \;G_{\;i,N-i}\;), \ \ 
G_{i,k} = \frac{\partial_s^i}{i!}\;\frac{\partial_t^{k}}{k!}\;G(s,t)
\end{equation}
For $i+k=N$ the integral for $G_{i,k}$ becomes logarithmically divergent and $L.P. (G_{i,k})$ is a constant which we  calculate using the sector decomposition technique. It turns out that it is more convenient to calculate the graph $\tilde G_{i,k}$ with massive lines and to put  $s=t=0$. The
 leading pole of such a graph is exactly the same as for the massless graph
 \begin{equation}
L.P. \left(\tilde G_{i,k}|_{s,t=0}\right) \equiv L.P. (G_{i,k});
\label{gamma_tilde}
\end{equation}
 however, choosing all momenta to be zero one gets additional simplifications in the sector decomposition technique.

The standard form of the alpha-representation in the Euclidean space for a graph without the numerators is
\begin{equation}
\label{alpha1}
I(s,t,m_i) = \frac{(\pi)^{DL/2}}{\prod\limits_{i=1}^{n}\Gamma(\lambda_i)}\int_0^\infty d\alpha_1...d\alpha_n \frac{\prod\limits_{i=1}^{n}\alpha_i^{\lambda_i -1}}{U^{d/2}} e^{-V/U-\sum\limits_{j=1}^{n} m_j\alpha_j}
\end{equation}
where $\lambda_i$ are the powers of the propagators, the functions $U$ and $V$ are the polynomials of the alpha-parameters ($\alpha_i$) of the order $L$ and $L+1$, respectively. The polynomial $V$ linearly depends on the squared combinations of the external momenta of the graph. In our case 
\begin{equation}
\label{alpha_v}
V=s \cdot P_s(\alpha_i) +t \cdot P_t(\alpha_i) \;.
\end{equation}

\begin{figure}[ht]
\centering
\includegraphics[scale=0.40]{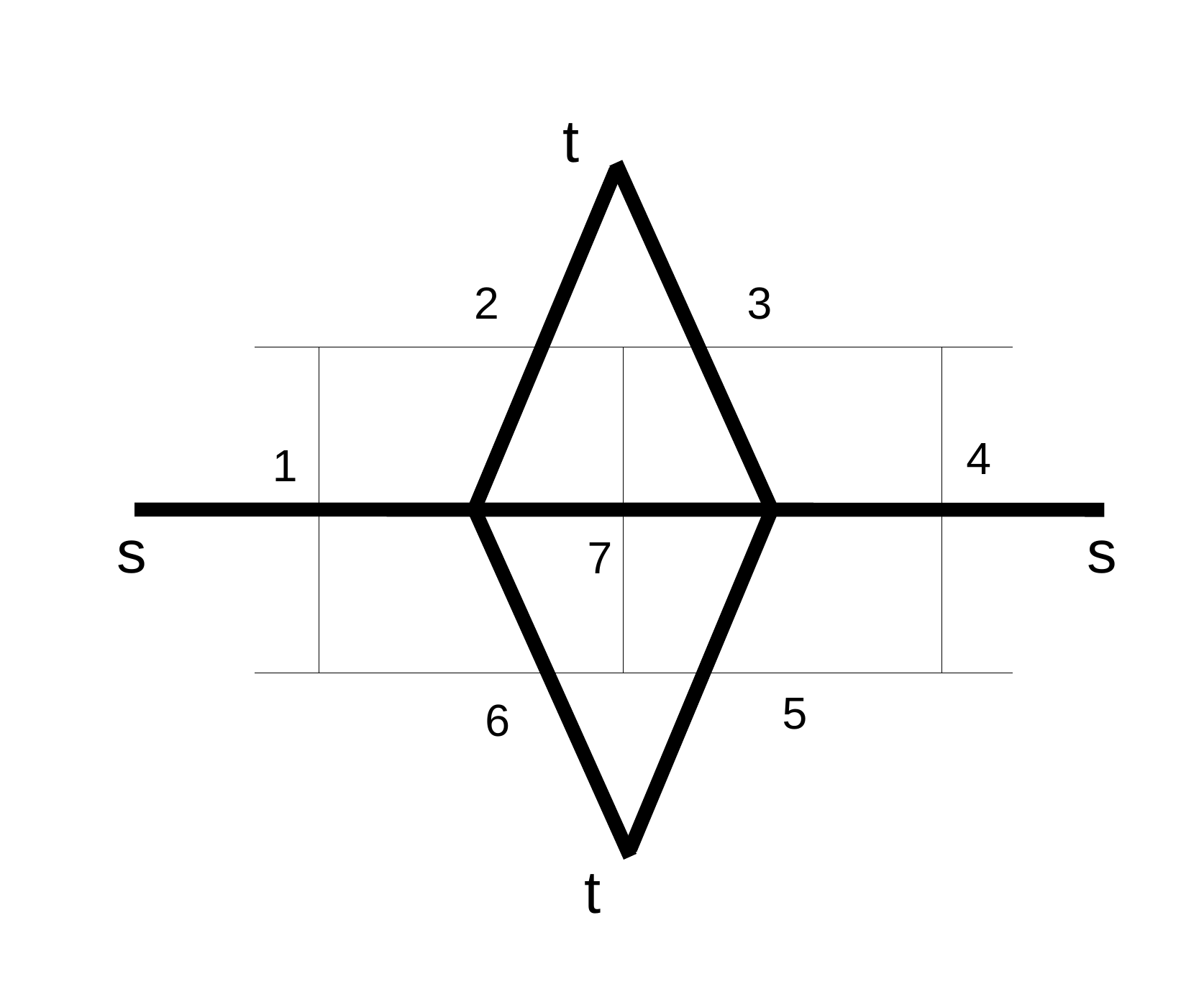}
\caption{The double box diagram and its dual graph}
\label{db_dual}
\end{figure}

\begin{figure}[ht]
\centering
\includegraphics[scale=0.4]{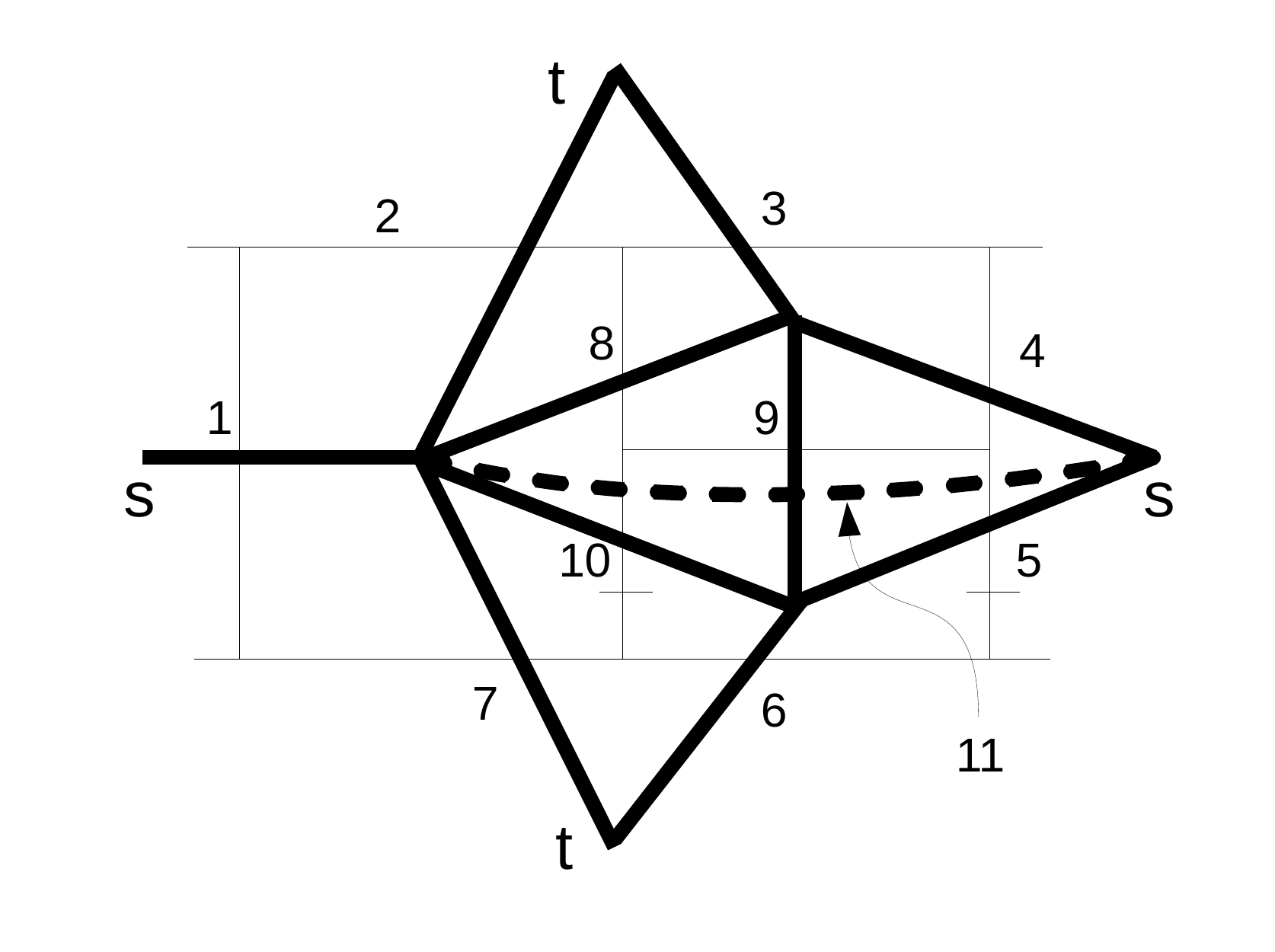}
\caption{The tennis-court diagram and its dual graph}\label{tc}
\end{figure}

In the general case (for the graphs with the numerators), the alpha-representation can be obtained from the generating function (for details see \cite{Zavialov}), which  allows one to construct the alpha-representation for any particular numerators. This procedure though straight\-forward is rather lengthy. Instead, 
since we have very specific numerators, we  construct the generating function in terms of  the  so-called \textit{dual graphs}. 
(Some examples are given in Figs.\ref{db_dual} and \ref{tc}).
 Note that this is always possible for the planar diagrams. 
In the dual representation the numerators can be expressed as additional lines with negative  powers of the propagators (see the dotted line in Fig.\ref{tc})).

The propagators with negative powers can be treated as the ordinary propagators but instead of integration over the alpha-parameter one should differentiate with respect to it and after that set it to zero
\begin{equation}
I(s,t,m_i) = \frac{(\pi)^{DL/2}}{\prod\limits_{i=1}^{n}\Gamma(\lambda_i)}\left(\left(\prod\limits_{i=n+1}^{n+k}(-\partial_{\alpha_i})^{\kappa_i}\right)\int_0^\infty \frac{d\alpha_1...d\alpha_n }{U^{d/2}} e^{-V/U-\sum\limits_{j=1}^{n} m_j\alpha_j}\right)\Big|_{\alpha_{n+1}=...=\alpha_{n+k}=0}
\end{equation}
Here the parameters $\alpha_{n+1},...,\alpha_{n+k}$ correspond to the numerators, and $\kappa_{n+1},...,\kappa_{n+k}$ are the powers of the corresponding numerators. The advantage of this approach with respect to the standard one is that we have only two polynomials $U$ and $V$ and no additional terms appear.

For the graphs without the numerators the monomials in $U$ are generated by the 1-trees and have a degree of $L$. The  momenta corresponding to the alpha-parameters in each monomial should be linearly independent (assuming that all external momenta and their combinations are zero). In terms of the dual graphs they should not form  loops. Each monomial in $V$ is generated by the 2-trees and contain the factor equal to the squared momenta flowing from the one part of the 2-tree to the other. 

Note that for the polynomial $U$ in the dual representation one has to treat all external vertices as identical (equivalent to setting all combinations of external momenta to zero) and avoid the combinations with chains that contain two external vertices.
The polynomial $V$ is constructed in the same way, but monomials must contain exactly one chain between external vertices,  and is multiplied by the   combination of external momenta squared flowing through the lines that belong to this chain. Actually, this chain works like a cut in the standard (not dual) representation and the momentum factor is the momentum flowing through this cut (this is exactly what 2-trees do). Since in  our case we have only momenta $s$ and $t$  in $V$, this means that for $P_s$
we need the chains that connect the $t$-vertices and for $P_t$ the  chains that connect the $s$-vertices (see example below). 

To illustrate the equivalence of the dual graphs approach with the usual 1- and 2- trees, we consider first  a simple example without the numerators (see Fig.\ref{db_dual}). 

To construct the polynomial $U$ for this graph, we need to find all possible pairs of alpha parameters which do not form the loops in the dual graph while all external vertices (marked by $s$ and $t$) are considered as identical. One has the following combinations of two and three alpha-parameters which form the loops in the dual graph:
\begin{equation}
\{\alpha_1 \alpha_2, \;\alpha_1 \alpha_6, \;\alpha_3 \alpha_4, \;\alpha_3 \alpha_5, \;\alpha_3 \alpha_6, \; \alpha_4 \alpha_5\}.
\label{db_cons2}
\end{equation}
\begin{equation}
\{ \alpha_1\alpha_3\alpha_7,\;\alpha_1\alpha_4\alpha_7,\;\alpha_1\alpha_5\alpha_7,\;\alpha_2\alpha_3\alpha_7,\;\alpha_2\alpha_4\alpha_7,\;\alpha_4\alpha_6\alpha_7,\;\alpha_5\alpha_6\alpha_7,\;\alpha_2\alpha_5\alpha_7,\;\alpha_3\alpha_6\alpha_7\}.
\label{db_cons3}
\end{equation}
Since the polynomial $U$ has degree of 2 we take all possible pairs of alpha-parameters except for those listed in \eqref{db_cons2} and get:
\begin{equation}
\begin{split}
U =&\;  \alpha_1\alpha_3 + \alpha_1\alpha_4 + \alpha_1\alpha_5 +\alpha_1\alpha_7 + \alpha_2\alpha_3 +\alpha_2\alpha_4 + \alpha_2\alpha_5 +\alpha_2\alpha_7 +\alpha_3\alpha_6 +\alpha_3\alpha_7 +\\
\,+&\alpha_4\alpha_6 +\alpha_4\alpha_7 + \alpha_5\alpha_6 + \alpha_5\alpha_7 +\alpha_6\alpha_7
\end{split}
\label{db_u}
\end{equation}
It is easy to check that this polynomial is exactly the one constructed using the 1-trees. The excluded combinations of alpha-parameters \eqref{db_cons2} are the ones for which the rest  of the graph does not form the 1-tree. 

To get the polynomial $V$ \eqref{alpha_v}, one needs to find $P_s$ and $P_t$. $P_s$  consists of the combinations of three alpha-parameters which contain the chain of lines connecting the $t$-vertices and does not form any other loops in the dual graph (the same for $P_t$  but the chain should connect the $s$-vertices). There are only 4 chains that connect the $t$-vertices:
\begin{equation}
\{\alpha_2\alpha_6,\; \alpha_3\alpha_5,\; \alpha_2\alpha_5\alpha_7,\; \alpha_3\alpha_6\alpha_7\}.  
\label{db_chain_t}
\end{equation}
Each monomial in $P_s$ is the combination of three alpha-parameters which contain the chains from \eqref{db_chain_t} and do not form any additional loops \eqref{db_cons2},\eqref{db_cons3}. Thus, for $P_s$ we have:
\begin{equation}
P_s = \bm{\alpha_2\alpha_6} (\alpha_3 + \alpha_4+\alpha_5 + \alpha_7) + \bm{\alpha_3\alpha_5}(\alpha_1+\alpha_2+\alpha_6+\alpha_7) +  \bm{\alpha_2\alpha_5\alpha_7} + \bm{\alpha_3\alpha_6\alpha_7} .
\end{equation}
For $P_t$ there is only one chain ($\{\alpha_1\alpha_4\alpha_7\}$) that connects the $s$-vertices, so $P_t$ has form:
\begin{equation}
P_t = \bm{\alpha_1\alpha_4\alpha_7}
\end{equation}
This concludes the construction of all ingredients needed for the alpha-representation of the double box graph.

Consider now the "tennis-court" diagram (Fig. \ref{tc}). The dashed line ($\alpha_{11}$) in the figure corresponds to the numerator and in terms of the dual graph is treated as a normal line with the propagator in the negative power. The polynomial $U$ contains the monomials with all possible combinations of 3 alpha-parameters except for the ones which form the loops in the dual graph. For this graph, the  loops are formed by the following combinations of alpha-parameters: $\alpha_1\alpha_2$, $\alpha_3\alpha_4$, $\alpha_5\alpha_6$, $\alpha_1\alpha_7$, $\alpha_2\alpha_7$, $\alpha_1\alpha_{11}$, $\alpha_2\alpha_{11}$, $\alpha_7\alpha_{11}$ and  $\alpha_2\alpha_3\alpha_8$, $\alpha_4\alpha_8\alpha_{11}$, $\alpha_4\alpha_5\alpha_{11}$, etc. (totally there are 21 combinations which contain 3 parameters and 22 combinations with 4 parameters). This results in 80 terms for $U$.
The polynomial $P_s$ contains the chains that connect the  t-vertices and contain no other loops, i.e., $\alpha_2\alpha_3\alpha_9\alpha_7$, etc., (totally 32 terms). For $P_t$ there are 25 terms.

Thus, this  diagram in the alpha-representation looks like
\begin{equation}
I_{t.c.}(s,t,m_i) = (\pi)^{3D/2}\left((-\partial_{\alpha_{11}})\int_0^\infty \frac{d\alpha_1...d\alpha_{10} }{U^{d/2}} e^{-(s P_s+ t P_t)/U-\sum\limits_{j=1}^{10} m_j\alpha_j}\right)\Big|_{\alpha_{11}=0}
\label{tc_fr}
\end{equation}
In the case of $D=6$, this diagram diverges logarithmically, and we can set both $s$ and $t$ to zero and calculate the integral
\begin{equation}
\tilde{G}_{0,0}^{(D=6)}(s=0,t=0,m_i) = (\pi)^{3D/2}\left((-\partial_{\alpha_{11}})\int_0^\infty \frac{d\alpha_1...d\alpha_{10} }{U^{d/2}} e^{-\sum\limits_{j=1}^{10} m_j\alpha_j}\right)\Big|_{\alpha_{11}=0}
\label{tc6}
\end{equation}

In the $D=8$ case, the graph has degree of divergence equal to 6. Hence,  one has four contributions proportional to $s^3$, $s^2t$, $s t^2$ and $t^3$. Taking the corresponding derivatives one has
\begin{equation}
\tilde{G}_{i,3-i}^{(D=8)}(s=0,t=0,m_i) = (\pi)^{3D/2}\!\left(\!\!(-\partial_{\alpha_{11}})\!\int_0^\infty \!\!\frac{d\alpha_1...d\alpha_{10} (-P_s)^i (-P_t)^{3-i} }{U^{d/2+3}} e^{-\sum\limits_{j=1}^{10} m_j\alpha_j}\!\right)\!\Big|_{\alpha_{11}=0}
\label{tc8}
\end{equation}
Here the extra multiplier $(-P_s)^i (-P_t)^{3-i} /{U^{3}}$ originates from differentiation of \eqref{tc_fr} with respect to $s$ and $t$.

For the integrals like \eqref{tc6},\eqref{tc8} the leading pole can be extracted using the sector decomposition method. To do this, we adopt the Speer-like strategy \cite{strategyS} for the dual graphs. For the tennis court diagram this strategy produces 390 sectors, for $D=6$ all the sectors contribute to the  leading pole, but for $D=8$ the leading pole is present only in 144 sectors out of 390.

We performed the numerical evaluation of the leading pole for the  diagrams up to 4 loops in $D=6$ and $D=8$, and up to 3 loops in $D=10$. The obtained results  are in  good agreement with the analytical values, as can be seen from Table \ref{num_d6_d8}. The overall computational time is about 300 hours at the 140 core cluster. 

\begin{table}[!htb]
    \begin{minipage}{.48\linewidth}
      \label{num_d6}
      \centering
        \begin{tabular}{|c|c|c|c|}\hline
            graph & term & numerical & exact\\
            \hline
            $I_1^{(4)}$ & $s^0\,t^0$& 0 & 0 \\
            \hline
            \multirow{3}{*}{$I_2^{(4)}$} & \multirow{3}{*}{$s^0\,t^0$}& \multirow{3}{*}{0.0416652(17)}& \multirow{3}{*}{1/24} \\
             & & & \\
             & & & \\
            \hline
            \multirow{3}{*}{$I_3^{(4)}$} & \multirow{3}{*}{$s^0\,t^0$}&\multirow{3}{*}{0.0208328(7)} & \multirow{3}{*}{1/48}\\
             & & & \\
             & & & \\ \hline
        \end{tabular}
    \end{minipage}%
    ~~
    \begin{minipage}{.48\linewidth}
      \centering
        \label{num_d8}
        \begin{tabular}{|c|c|c|c|}\hline
            graph & term & numerical & exact\\
            \hline
            $I_1^{(4)}$ & $s^3$ & $-209.997(5)$& $-210$ \\
            \hline
            \multirow{3}{*}{$I_2^{(4)}$} & $s^4$& -6.6661(10) & -20/3\\
                        & $s^3t$& 0.888900(24) & 8/9\\
                        & $s^2t^2$& -0.1111105(7) & -1/9\\
            \hline
            \multirow{3}{*}{$I_3^{(4)}$} & $s^4$& -20.4765(8) & -430/21 \\
                        & $s^3t$& 0.444420(25) & 4/9\\
                        & $s^2t^2$& -0.0555541(10) & -1/18\\

            \hline
        \end{tabular}
    \end{minipage}
    \caption{The numerical values for some sample 4-loop graphs in $D=6$ (left) and $D=8$ (right). The values shown in the right table are multiplied by $6!5!4!3!$. }
    \label{num_d6_d8}
\end{table}
\newpage
\section{Summary of the leading pole evaluation in various dimensions}

We summarize here the results of  calculation of the leading poles in various dimensions.
 
\subsection{$D=6$ $\mathcal{N}=(1,1)$ SYM}
Summarizing one has for the leading poles (L.P.) \cite{BKV}
\begin{equation}
L.P.=2st g^4\left[ g^2  \frac{s+t}{6\epsilon}+g^4  \frac{s^2+st+t^2}{36\epsilon^2}+g^6  \frac{s^3+\frac 25 s^2t+\frac 25 st^2+t^3}{216\epsilon^3}\right]
\end{equation}
The leading powers of $s$ ant $t$ remind the geometrical progression while for the mixed ones there are too few terms to make any guess. If taking the geometrical progression seriously, one gets
\begin{equation}
\sum_{n=1}^\infty  \left(\frac{g^2 s}{6\epsilon}\right)^n = \frac{\frac{g^2s}{6\epsilon}}{1-\frac{g^2s}{6\epsilon}},
\label{geom}
\end{equation}
which looks precisely like the D=4 Yang-Mills theory with the replacement $g^2\to g^s s$ and in the limit when 
$\epsilon \to +0$ tends to $-1$ when $s<0$ and to $\infty$ when $s>0$.  
A natural question arises whether one can prove eq.(\ref{geom}) to be correct.  For this purpose consider the sequence of diagrams appearing in the loop expansion (\ref{expan}).

\begin{figure}[ht]
\begin{center}
\includegraphics[width=0.95\textwidth]{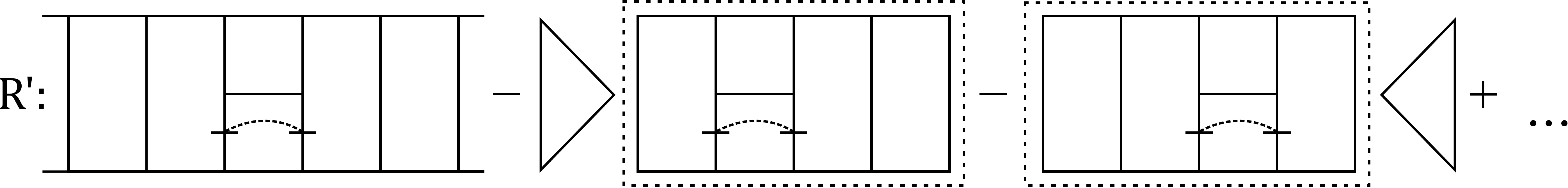}
\caption{${\cal R}'$-operation for the ladder-court graph. Shown are the one loop surviving graphs only. The dotted line denotes the contracted graph for which the $K{\cal R}'$ has to be taken. \label{Rop}}
\end{center}
\end{figure}

We start with the infinite sequence of diagrams originating from the graph $I_2^{(3)}$ by adding the boxes to the left and to the right. This gives us the diagrams $I_2^{(4)}, I_3^{(4)}, I_1^{(5)}, I_2^{(5)}$, etc. Performing the ${\cal R}'$-operation and looking for the surviving one loop diagrams one can notice that they can stand either on the left or right edge of the diagram or in the middle. But the tennis court subgraph, $I_2^{(3)}$,  can be present only once since it is a three-loop block and if it stands twice the order of the pole drops by two. And the diagrams containing only boxes do not diverge. Hence one is left with one option: the one loop graph  stands at the edge. This is shown in Fig.\ref{Rop}.

Consider  first the case  when the tennis court is situated at the edge and the boxes are added to one side. Then, since the triangle graph itself is equal to  $1/2\epsilon$, the $K{\cal R}' G_{n-1}=(-1)^n A_{n-1}$ and taking into account eq.(\ref{rel}) one gets
\beq
n A_n=-\frac 12 A_{n-1},\label{one}
\eeq
where $n$ is the total number of loops. It has a solution
\beq
A_n=\frac{(-1)^n}{2^n}\frac{c}{n!},\label{ones}
\eeq
where the constant $c$ can be  found from the $n=3$ case and is equal to 8.

For the general case, when the boxes are added from both sides one has to take the sum of all diagrams with k boxes to the left and n-2-k boxes to the right and sum over k from 0 to n-2. The sum obeys the equation
\beq
n \Sigma_n=-\Sigma_{n-1}\label{two}
\eeq
with the solution
\beq
\Sigma_n=(-1)^{n}\frac{c}{n!}, \ \ \ c= 2. \label{twos}
\eeq
One can check that the corresponding diagrams $I_2^{(4)}, I_3^{(4)}, I_1^{(5)}, I_2^{(5)}$ are reproduced by eqs.(\ref{ones},\ref{twos}).

Note that factorial comes inevitably due to the linear nature of eqs.(\ref{one},\ref{two}). It reflects the fact that the boxes themselves are finite. Therefore, this type of behavior will take place for all series of diagrams in this theory, in particular the ones that start with the diagram $I_5^{(4)}$  and the following ones. In any new order of PT the new series starts.

Knowing the n-th order coefficient one may sum all of them and get
\beq
\sum_{n=3}^\infty  \left(\frac{g^2 s}{\epsilon}\right)^{n-2} \frac{2}{n!} = \frac{2\epsilon^2}{g^4s^2}\left(
e^{\frac{g^2 s}{\epsilon}}-1-\frac{g^2 s}{\epsilon}-\frac 12(\frac{g^2 s}{\epsilon})^2\right) \to  \left\{ \begin{array}{ll} -1& s<0\\ \infty & s>0 \end{array}\right.    \ \ \ \mbox{when} \ \ \epsilon \to +0 \label{exp}
\eeq
One can see that we get the same result in asymptotic as in eq.(\ref{geom}), but we have taken into account only one sequence of diagrams which give the main contribution to the coefficients.

One can continue this procedure and sum the diagrams of the next series which starts with the 4-loop diagram $I_5^{(4)}$. The difference here is that $I_5^{(4)}$ is not a constant but is proportional to $t-s$. As for $s$, it stands outside the integration and is not changed but $t$ is replaced by $t'$ corresponding to the contracted diagram in analogy with Fig.\ref{Rop}
and has to be integrated over triangle giving both $s$ and $t$. Thus, we have two relations, one proportional to $s$, and the other to $t$. Considering the sequence of diagrams where the boxes are added from one side, one gets the relations
\beq
n A^{t}_n=-\frac 16 A^{t}_{n-1},  \ \ \ \ \  n A^{s}_n=-\frac 12 A^{s}_{n-1}+\frac 16 A^t_{n-1}. \label{onet}
\eeq
Solution to these relations is
\beq
A^t_n=\frac{(-1)^n}{6^{n-3}}\frac{1}{n!}, \ \ \ \ \  A^s_n=\frac{1}{2}\frac{(-1)^n}{6^{n-3}}\frac{1}{n!}-\frac 12 \frac{(-1)^n}{2^{n-3}}\frac{1}{n!} \label{onest}
\eeq
One can check that eq.(\ref{onest}) is valid for the diagrams $I_5^{(4)}$ and $I_4^{(5)}$. Summing up the diagrams where the boxes are added from both sides, similarly to the previous case, one has
\beq
n \Sigma^{t}_n=-\frac 13 \Sigma^{t}_{n-1},  \ \ \ \ \  n \Sigma^{s}_n= - \Sigma^{s}_{n-1}+\frac 13 \Sigma^t_{n-1} \label{twot}
\eeq
with the solution
\beq
\Sigma^t_n=\frac{(-1)^n}{3^{n-3}}\frac{1}{n!}, \ \ \ \ \  \Sigma^s_n=\frac 12\frac{(-1)^n}{3^{n-3}}\frac{1}{n!}-\frac 12 (-1)^n\frac{1}{n!} \label{twost}
\eeq
These relations reproduce the diagrams $I_5^{(4)},I_3^{(5)}$ and $I_4^{(5)}$.

Having these coefficients one can again calculate the whole  series
\beqa
&&3  \frac ts \sum_{n=4}^\infty   \left(\frac{g^2 s}{3\epsilon}\right)^{n-2} \frac{1}{n!} =\frac ts\frac{27\epsilon^2}{g^4s^2}\left(
e^{\frac{g^2 s}{3\epsilon}}-1-\frac{g^2 s}{3\epsilon}-\frac 12(\frac{g^2 s}{3\epsilon})^2-\frac 16(\frac{g^2 s}{3\epsilon})^3\right)\nonumber \\
&&  \to  \left\{ \begin{array}{ll} -\frac ts\frac{3}{2}[1+\frac{1}{3} (\frac{g^2 s}{3\epsilon})]& s<0\\ \infty & s>0 \end{array}\right.    \ \ \ \mbox{when} \ \ \epsilon \to +0 , \label{expt}
\eeqa

\beqa
 &&\frac 12 \sum_{n=4}^\infty  \left(\frac{g^2 s}{\epsilon}\right)^{n-2} \frac{1}{n!} ((\frac{27} {3^n}-1)=\frac{\epsilon^2}{2g^4s^2}\left[27 \left(e^{\frac{-g^2s}{3\epsilon}}-1-\frac{g^2 s}{3\epsilon}-\frac 12(\frac{g^2 s}{3\epsilon})^2-\frac 16(\frac{g^2 s}{3\epsilon})^3\right)\right. \nonumber \\
 &&\left. -\left(e^{\frac{-g^2s}{\epsilon}}-1-\frac{g^2 s}{\epsilon}-\frac 12(\frac{g^2 s}{\epsilon})^2-\frac 16(\frac{g^2 s}{\epsilon})^3\right)\right] \nonumber \\
 && \to  \left\{ \begin{array}{ll}   -\frac{3}{4}[1+\frac{1}{3} (\frac{g^2 s}{3\epsilon})]+\frac 14[1+\frac 13(\frac{g^2 s}{\epsilon})]=-\frac{1}{2},& s<0\\ \infty & s>0 \end{array}\right.  
  \ \ \ \mbox{when} \ \ \epsilon \to +0. \label{exps}
\eeqa

Thus, we see that in the limit $\epsilon\to +0$  when $s<0$ the first series (\ref{exp}) tends to a constant and the second to a constant plus the first pole (\ref{expt},\ref{exps}). Obviously, the third series, which starts from 5 loops, will tend to a constant, the first pole, the second pole and so on. This new series has to be summed again. It has a feature common to all the sequences, namely it falls as $1/n!$. At the same time the number of diagrams in each order is expected to be of the order of $n! n^b a^n$.  For $s>0$ all the series diverge when
$\epsilon \to +0$.

The same is true to the diagrams in the $t$-channel with the obvious replacement $s\leftrightarrow t$. In the next section, we will show how these results can be  promoted  for the general case.

 \subsection{$D=8$ $\mathcal{N}=1$ SYM}

Summarizing one has for the leading poles 
\beqa
L.P.&=&-st \left[ g^2  \frac{1}{3!\epsilon}+g^4  \frac{s^2+t^2}{3!4!\epsilon^2}+g^6 \frac{4}{3} \frac{15s^4-s^3t+s^2t^2-st^3+15 t^4}{3!4!5!\epsilon^3}\right. \\
&+&\left. g^8 \frac{1}{63} \frac{16770s^6-536s^5t+412s^4t^2-384s^3t^3+412s^2t^4-536st^5+16770t^6}{3!4!5!6!\epsilon^4}\right].\nonumber
\eeqa
 This expression does not look simple even for the leading powers of $s$ or $t$  though numerically one almost has a geometrical progression with slightly rising coefficients.
 
\begin{figure}[htb]
\begin{center}
\includegraphics[width=0.9\textwidth]{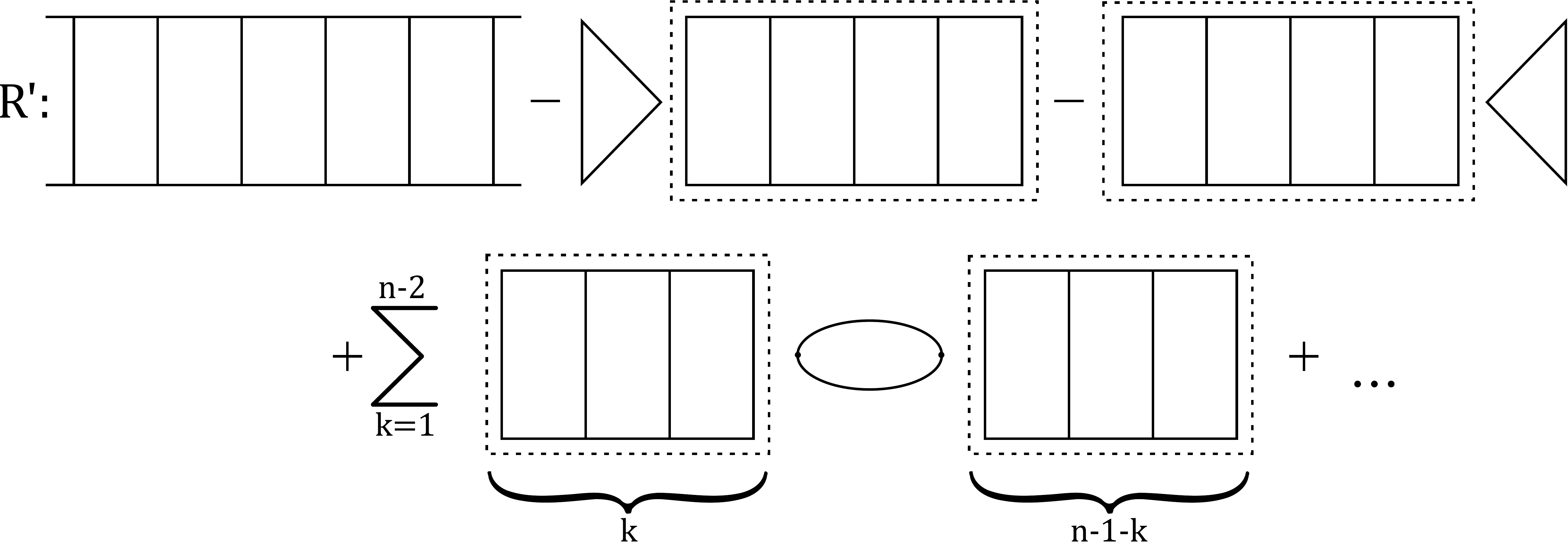}
\caption{${\cal R}'$-operation for the ladder type graph. Shown are the 1-loop surviving graphs only\label{Rop8}}
\end{center}
\end{figure}

 Consider now the infinite sequence of horizontal boxes and apply the ${\cal R}'$-operation. The difference from the previous case  is that now the box diagram is divergent, and performing the ${\cal R}'$-operation and looking for the surviving one loop diagrams one  has to consider the diagram in the middle as well.  Therefore, the graphical form in Fig. \ref{Rop} changes. It is shown in Fig.\ref{Rop8}.

Since the triangle diagram in D=8 is equal to $-1/4!/\epsilon$ and the bubble one to $2/5!/\epsilon$, this gives us the recurrence relation for the leading pole terms
\beq
n A_n=-\frac{2}{4!} A_{n-1}+\frac{2}{5!}\sum_{k=1}^{n-2}A_kA_{n-1-k}, \ \ \  n\geq 3 \label{one8}
\eeq
with $A_1=1/3!$. Starting from this value one can calculate any $A_n$ though the explicit solution is not straightforward. However, since we actually need the sum of the coefficients we apply the summation multiplying both sides of eq.(\ref{one8}) by $(-z)^{n-1}$, $z$ being $g^2s^2/\epsilon$
\beq
\sum_{n=3}^\infty  n A_n (-z)^{n-1}=-\frac{2}{4!}\sum_{n=3}^\infty A_{n-1} (-z)^{n-1}+\frac{2}{5!}\sum_{n=3}^\infty 
\sum_{k=1}^{n-2}A_k (-z)^k A_{n-1-k} (-z)^{n-1-k}.
\eeq
Denoting now the sum $\sum_{n=m}^\infty A_n (-z)^n$ by $\Sigma_m$ and performing the interchange of the order of summation in the nonlinear term we get
\beq
-\frac{d}{dz}\Sigma_3=-\frac{2}{4!}\Sigma_2+\frac{2}{5!}\Sigma_1\Sigma_1.
\eeq
Having in mind that 
$$\Sigma_3=\Sigma_1+A_1 z-A_2 z^2, \ \ \Sigma_2=\Sigma_1+A_1 z, \ \ A_1=\frac{1}{3!},\ A_2=-\frac{1}{3!4!},$$
one finally gets the equation for $\Sigma\equiv\Sigma_1$
\beq
\Sigma'=-\frac{1}{3!}+\frac{2}{4!}\Sigma-\frac{2}{5!}\Sigma^2.
\eeq
Solution to this equation is
\beq
\Sigma(z)=-\sqrt{5/3} \frac{4 \tan[z/(8 \sqrt{15})]}{1 - \tan[z/(8 \sqrt{15})] \sqrt{5/3}},\label{sol}
\eeq
The expansion of $\tan z$ contains the Bernuli numbers 
$$ \tan z = \sum_{n=1}^\infty (-1)^{n-1}\frac{2^{2n}(2^{2n}-1)B_{2n}}{(2n)!}z^{2n-1}.$$
Being substituted into eq.(\ref{sol}) it gives
\beq \Sigma(z)=-(z/6+z^2/144+z^3/2880+7z^4/414720 + \dots) \eeq
Remind that here  $z = \frac{g^2 s^2}{\epsilon}$.  
This series reproduces the diagrams $I_1^{(1)}, I_1^{(2)}, I_1^{(3)}$ and $I_1^{(4)}$ which give the main contribution to the amplitude PT series.

The function $\Sigma$  given by eq.(\ref{sol})  has an infinite sequence of simple poles and thus has no limit when $\epsilon \to 0$. This is similar to the geometrical progression with non-alternating series.

The next sequence of diagrams comes from the tennis-court one supplemented by boxes from both ends.
Here one again has the nonlinear terms like in eq.(\ref{one8}) and the set of three sums proportional to $s^2$, $st$ and $t^2$ like in eq.(\ref{twot}). This looks more complicated but is simplified by the fact that the tennis court diagram appears only ones, so equations are in fact linear and generally look like
\beq
\Sigma_{court}' = c-\Sigma_{court}+\Sigma_{court}\Sigma_{box}
\eeq
with $\Sigma_{box}$ given by eq.(\ref{sol}) above.

Since $\Sigma_{box}$  has a singular behavior when $\epsilon \to 0$, so does $\Sigma_{court}$.  Moreover, 
for any finite number of sequences  of this kind one will always have a singular behavior. Only when one has an infinite number of them, which is actually our case, one may avoid this singularity.

 \subsection{$D=10$ $\mathcal{N}=1$ SYM}

Summarizing one has for the leading poles 
\beqa
L.P.&=&-st \left[ g^2  \frac{s+t}{5!\epsilon}+g^4  \frac{8s^4+2s^3t+2st^3+8t^4}{5!7!\epsilon^2}\right. \\
&+&\left. g^6 \frac{2(2095s^7+115s^6t+33s^5t^2-11s^4t^3-11s^3t^4+33s^2t^5+115st^6+2095t^7)}{5!7!7!45\epsilon^3}\right. \nonumber \\
&+& \left. g^8\frac{32(211218880s^{10}+753490s^9t-1395096s^8t^2+1125763s^7t^3-916916s^6t^4}{13!7!7!5! 5\epsilon^4}\right. \nonumber \\
&&\hspace{-1.5cm} \left.\frac{+843630s^5t^5-916916s^4t^6+1125763s^3t^7-1395096s^2t^8+753490st^9+211218880t^{10})}{13!7!7!5! 5\epsilon^4}
\right].\nonumber
\eeqa

One can construct the recurrence relations here as well. For the box type diagrams one has a relation similar to the D=8 case but since the one loop box has the numerator $(s+t)$ one has two separate expressions like in eq.(\ref{onet}).  At the same time they are nonlinear like eq.(\ref{one8}). One has
\beqa
n A^{t}_n&=&-2\frac{2}{7!}A^{t}_{n-1}+\frac{1}{3 \cdot7!}\sum_{k=1}^{n-2}A^t_kA^t_{n-1-k}, \\ 
 n A^{s}_n&=&-2\left[\frac{1}{3 \cdot 5!}A^{s}_{n-1}-\frac{6}{7!} A^t_{n-1}\right]\\
 &+&\frac{3}{7!}\sum_{k=1}^{n-2}\left(2A^s_kA^s_{n-1-k}-A^s_kA^t_{n-1-k}-A^t_kA^s_{n-1-k}+\frac 59 A^t_kA^t_{n-1-k}\right)  \nonumber 
\eeqa
with $A_1^s=A_1^t=1/5!$. These equations also reproduce all the ladder type diagrams calculated above.

\section{All loop recursive equations}

It is possible to construct the recursive relations for all the diagrams. They actually follow from the analysis
of the series and the way how the diagrams are constructed and enter the ${\cal R}'$-operation. 

\begin{figure}[htb]
\begin{center}
\includegraphics[width=0.9\textwidth]{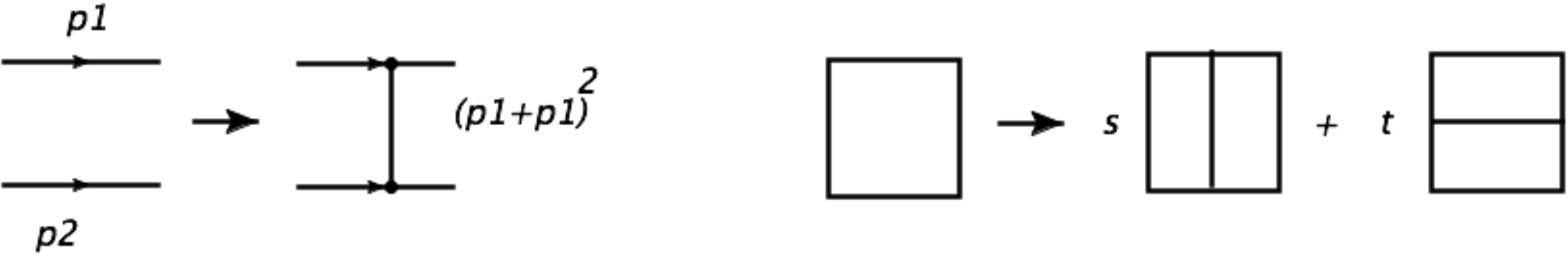}
\caption{The "rung rule" (left) and its application for the box diagram (right) \label{rung}}
\end{center}
\end{figure}

Indeed, one particular way to obtain the dual conformal invariant diagram of the n-th order from the (n-1)-order is to  follow the so-called "rung rule" \cite{RungRuler,BDS}, which states that one has to take the diagram of the (n-1)-th order and insert a line connecting each pair of the neighboring lines multiplying by a factor equal to the square of the total momentum (see Fig.\ref{rung} left). This is true for all the dual conformal invariant diagrams with triple vertices, which are proportional to the common factor $st$, and does not include the diagrams with numerators without this factor and the diagrams with  quartic vertices \cite{DixonBDS45loops}. However, the latter ones do not contain the leading poles.

To demonstrate how it works, let us take the simplest box diagram.  The application of the rung rule reproduces the horizontal and vertical double boxes with appropriate coefficients (see Fig.\ref{rung} right). This way, starting from a single box and applying the rung rule one adds boxes to the left and right and to the top and bottom, thus creating new $s$-channel and $t$-channel diagrams. 

What is important, to get the leading pole, the diagram should contain the maximally divergent subgraphs which being shirked to a point leave the one loop divergent graph. Integrating over this graph gives the desired leading pole. The topology of all the graphs constructed via the rung rule leaves just one possibility: the maximally divergent graph should have a box at the left (right) or top (bottom) edge. All the other graphs either do not contain the maximally divergent subgraphs or, if they do, their shrinking to a point leaves the one loop subgraph, which contains a single leg with light like momenta and hence is equal to zero. This property singles out the set of maximally divergent graphs.

Consider now the ${\cal R}'$-operation for this set of graphs. For the the $s-$channel type diagram it has the form graphically presented in Figs.\ref{Rop} and \ref{Rop8} where now the (n-1)-th order diagrams in dotted boxes contain  both the $s$-channel and $t$-channel contributions. Here we use the above mentioned property that the diagram of interest always contains the box either on the left or on the right edge.
The ${\cal K R}'G_{n-1}$ is a polynomial in $s$ and $t$; however, $s$ is a common factor while $t$ for $G_{n-1}$ contains the integration  momentum over the last loop. Substituting the explicit form of $s$ and $t$ and integrating over the triangle by introducing the Feynman parameters one gets the desired recursive relation.

\subsection{$D=6$ $\mathcal{N}=(1,1)$ SYM}
In the case of  $D=6$ it looks like
\beq
nS_n(s,t)=-2 s \int_0^1 dx \int_0^x dy  \ (S_{n-1}(s,t')+T_{n-1}(s,t')), \ \ \ \     n\geq 4\label{eq}
\eeq
where $t'= t x+u y$, $u=-t-s$, and   $S_3=-s/3,\ T_3=-t/3$.
Here we denote by  $S_n(s,t)$ and  $T_n(s,t)$  the sum of all contributions  in the  $n$-th order of PT in $s$ and  $t$ channels, respectively.  The same recursive relation is valid for the $t$-channel  diagrams with the obvious replacement $s\leftrightarrow t$. Due to  the $s-t$ symmetry of the amplitude, one should have $T_n(s,t)=S_n(t,s)$.  
The coefficient $A_n(s,t)$ of the $n$'th order pole is the sum
$$
A_n(s,t)=S_n(s,t)+T_n(s,t).
$$

Eqs. (\ref{eq})  reproduces all the diagrams calculated above and may serve as a generating function for the diagrams in all orders.  The first few terms are
\begin{eqnarray}
S_4(s,t)&=&+\frac{2s^2+st}{36},\nonumber\\
S_5(s,t)&=&-\frac{5s^3 + s^2 t + st^2}{540},\\
S_6(s,t)&=&+\frac{25 s^4 + 5 s^3 t - s^2 t^2 + 3 s t^3}{19440}\nonumber.
\end{eqnarray}
The Mathematica file with a simple
code, which generates the $S_n$ and $T_n$ polynomials up to given order is available.

Since we know now all the leading pole contributions in all orders of PT, it is tempting to sum them over.
However, explicit solution of the recursive relation (\ref{eq}) is problematic. Instead, we proceed in the following way: we multiply both sides of eq.(\ref{eq}) by $(-z)^{n-1}$ and sum over $n$ from 4 to $\infty$, $z$ being $g^2/\epsilon$.
Then on the left hand side one has a derivative
\beq
\frac{d}{dz}\Sigma_4(s,t,z)=2s  \int_0^1 dx \int_0^x dy\ (\Sigma_3(s,t',z)+\Sigma_3(t',s,z))|_{t'=xt+yu}. \label{s6}
\eeq
Using the fact that $\Sigma_4(s,t,z)=\Sigma_3(s,t,z)+S_3(s,t) z^3$ one gets the equation
\beq
\frac{d}{dz}\Sigma_3(s,t,z)=3S_3 z^2+2s \int_0^1 dx \int_0^x dy\ (\Sigma_3(s,t',z)+\Sigma_3(t',s,z))|_{t'=xt+yu}.\label{eqsum}
\eeq
Since the first divergence appears only in the third loop order the function of interest is $\Sigma(s,t,z)=\sum _{n=3}^\infty (-z)^{n-2}S_n=z^{-2}\Sigma_3(s,t,z)$. Substituting it into eq.(\ref{eqsum}) and taking into account that $S_3(s,t)=-s/3$ one gets the final equation for the sum of PT series
 \beq
 \frac{d}{dz}\Sigma(s,t,z)=s-\frac{2}{z}\Sigma(s,t,z)+2s \int_0^1 dx \int_0^x dy\ (\Sigma(s,t',z)+\Sigma(t',s,z))|_{t'=xt+yu}.
\label{eqr}
\eeq
One has the same equation in the  $t$-channel
 \beq
 \frac{d}{dz}\Sigma(t,s,z)=t-\frac{2}{z}\Sigma(t,s,z)+2t \int_0^1 dx \int_0^x dy\ (\Sigma(s',t,z)+\Sigma(t,s',z))|_{s'=xs+yu}.
\eeq
Summing them up one gets for the total  sum $\Sigma(s,t,z)+\Sigma(t,s,z)$ 
 \beqa
&&\frac{d}{dz}(\Sigma(s,t,z)+\Sigma(t,s,z))=(s+t)-\frac{2}{z}[\Sigma(s,t,z)+\Sigma(t,s,z)]\label{eqrsum}\nonumber\\&&+2s \int_0^1 dx \int_0^x dy\ [\Sigma(s,t',z)+\Sigma(t',s,z)]|_{t'=xt+yu}\\&&+2t \int_0^1 dx \int_0^x dy\ [\Sigma(s',t,z)+\Sigma(t,s',z)]|_{s'=xs+yu}.\nonumber
\eeqa

The behavior of the solution to this equation is defined by the fixed point, i.e. the zero of the right hand side.
As $z\to \infty$ one can neglect the second term and under the  assumption that the fixed point is a constant get the following conjecture: 
\beq  \Sigma(s,t,z)+\Sigma(t,s,z) =-1.\eeq.

Consider now how this fixed point is approached. The sign of the derivative is propor\-tional to $s+t=-u$. In the case when $u<0$ the derivative is positive above the fixed point and negative below it. So if the initial value of $\Sigma$ is above the fixed point, it will increase and if it is below it, it will decrease. This means that the fixed point is unstable. On the contrary, if $u>0$ the sign is changed and the fixed point is stable, the solution tends to it
as $z\to + \infty$ or $\epsilon\to +0$.  Therefore, the stability properties depend on the kinematic region.
For $u>0$ the fixed point is stable and the theory is finite in the limit $\epsilon\to +0$. 

The situation will be the same for other partial amplitudes. In the (s,u) channel the theory is finite if $t>0$ and in the (t,u) channel it is finite if $s>0$. Unfortunately, all three conditions are incompatible since $s+t+u=0$
and one can not have all of them to be positive.

\subsection{$D=8,10$ $\mathcal{N}=1$ SYM}
Consider now the cases of  $D$=8 and $D$=10.  Here according to Fig.\ref{Rop8}, one has an additional nonlinear term, therefore eq.(\ref{eq}) is modified. Note that for this last term when integrating over the loop on both sides one has functions of $s$ and $t$. Replacing $t$ by $t'$ one should have in mind that on the left $t'=(l-p_1)$ and on the right $t'=(l+p_4)$, where $l$ is the integration momentum. This means that while integration one gets the mixed terms like $g^{\mu\nu}p_1^\mu p_4^\nu$.  This can be taken into account and one gets an equation which looks the same way for $D$=8  and $D$=10. For $D$=8 one has
\beqa
&&nS_n(s,t)=-2 s^2 \int_0^1 dx \int_0^x dy\  y(1-x) \ (S_{n-1}(s,t')+T_{n-1}(s,t'))|_{t'=tx+yu}\label{eq8}\nonumber \\ &+&
s^4 \int_0^1\! dx \ x^2(1-x)^2 \sum_{k=1}^{n-2}  \sum_{p=0}^{2k-2} \frac{1}{p!(p+2)!} \
 \frac{d^p}{dt'^p}(S_{k}(s,t')+T_{k}(s,t')) \times \nonumber \\
&&\hspace{2cm}\times  \frac{d^p}{dt'^p}(S_{n-1-k}(s,t')+T_{n-1-k}(s,t'))|_{t'=-sx} \ (tsx(1-x))^p, 
\eeqa
where $S_1= \frac{1}{12},\ T_1=\frac{1}{12}$. Equation(\ref{eq8}) reproduces all the above calculated diagrams. The terms with the derivatives in the second term do not contribute
so far.

In the case of $D$=10 one gets analogously
\beqa
&&nS_n(s,t)=-s^3 \int_0^1 dx \int_0^x dy\  y^2(1-x)^2 \ (S_{n-1}(s,t')+T_{n-1}(s,t'))|_{t'=tx+yu}\label{eq10}\nonumber \\ &+&
s^5 \int_0^1\! dx \ x^3(1-x)^3 \sum_{k=1}^{n-2}  \sum_{p=0}^{3k-2} \frac{1}{p!(p+3)!} \
 \frac{d^p}{dt'^p}(S_{k}(s,t')+T_{k}(s,t')) \times \nonumber \\
&&\hspace{2cm}\times  \frac{d^p}{dt'^p}(S_{n-1-k}(s,t')+T_{n-1-k}(s,t'))|_{t'=-sx} \ (tsx(1-x))^p, 
\eeqa
where $S_1= \frac{s}{5!},\ T_1=\frac{t}{5!}$.
Here the terms with the derivatives work starting from 3 loops.

Equations (\ref{eq8}), (\ref{eq10}) can be summed the same way as in the $D$=6 case (\ref{s6}). Multiplying both sides by  $(-z)^{n-1}$ and summing up over $n$ from 3 to   $\infty$ one gets
\beqa
&&\frac{d}{dz}\Sigma_3(s,t,z)=2 s^2 \int_0^1 dx \int_0^x dy\  y(1-x)\ (\Sigma_2(s,t',z)+\Sigma_2(t',s,z)
|_{t'=tx+yu}\\
&&-s^4  \int_0^1\! dx \ x^2(1-x)^2 \sum_{p=0}^\infty \frac{1}{p!(p+2)!} (\frac{d^p}{dt'^p}(\Sigma_1(s,t',z)+\Sigma_1(t',s,z)|_{t'=-sx})^2 \ (tsx(1-x))^p. \nonumber
\eeqa
Using now that
$$ \Sigma_3(s,t,z)=\Sigma_1(s,t,z)^s-S_2(s,t) z^2+S_1(s,t)z, \  \Sigma_2(s,t,z)=\Sigma_1(s,t,z) +S_1(s,t)z, $$
$$\frac{d}{dz}\Sigma_3(s,t,z)= \frac{d}{dz}\Sigma_1(s,t,z)- 2S _ 2(s,t)z+S_1(s,t), \ \  2S_2(s,t)=2s^2\int (S_1(s,t')+S_1(t',s))$$
one gets
\beqa
&&\frac{d}{dz}\Sigma(s,t,z)=-\frac{1}{12}+2 s^2 \int_0^1 dx \int_0^x dy\  y(1-x)\ (\Sigma(s,t',z)+\Sigma(t',s,z))|_{t'=tx+yu}\\
&&-s^4  \int_0^1\! dx \ x^2(1-x)^2 \sum_{p=0}^\infty \frac{1}{p!(p+2)!} (\frac{d^p}{dt'^p}(\Sigma(s,t',z)+\Sigma(t',s,z))|_{t'=-sx})^2 \ (tsx(1-x))^p. \nonumber
\eeqa
And analogously for $D$=10
\beqa
&&\frac{d}{dz}\Sigma(s,t,z)=-\frac{s}{5!}+s^3 \int_0^1 dx \int_0^x dy\  y^2(1-x)^2\ (\Sigma(s,t',z)+\Sigma(t',s,z))|_{t'=tx+yu}\\
&&-s^5  \int_0^1\! dx \ x^3(1-x)^3 \sum_{p=0}^\infty \frac{1}{p!(p+3)!} (\frac{d^p}{dt'^p}(\Sigma(s,t',z)+\Sigma(t',s,z))|_{t'=-sx})^2 \ (tsx(1-x))^p. \nonumber
\eeqa
The same equation with the replacement $s \leftrightarrow t$ can be derived for $\Sigma(t,s,z)$.

\section{Conclusion}

Summarizing the obtained results one should admit that the calculation of the leading poles of the multiloop diagrams in nonrenormalizable theories is not a simple task contrary to the renormalizable theories where they are given by the renormalization group equations. Nevertheless, we succeeded in writing down the recursive equations which allow us to calculate the desired poles at any loop order. In a sense, these equations replace the RG ones for the nonrenormalizable case \cite{DI_nonerenorm_RG}. The difference is that in this case they are not algebraic but contain the integration  over  the Feynman parameters.

Summation of perturbation series for the leading poles is also more complicated though the qualitative behaviour resembles that of renormalizable theories with the obvious replacement $g^2\to g^2 s$ or
$g^2t$. To get the sum of the infinite series, we derived the differential equations and the task is reduced to
the problem of finding the fixed points and investigating their stability properties. 
This is more explicit in $D=6$ and looks more complicated in $D=8,10$ due to the nonlinearity of the equations.

In renormalizable theories the UV divergences either cancel in each order of PT like in the $\mathcal{N}=4$
$D$=4 SYM theory, or are absorbed into the renormalization of the couplings and fields. In the nonrenormalizable case one may hope  to get either the cancellation at each order (see the attempts in  $\mathcal{N}=8$ SUGRA in $D$=4) or the finite limit as a result of summation of the infinite PT series  (like in D=4 QED where one has a zero charge behaviour).  
We have demonstrated that in higher dimensional maximally supersymmetric theories one may  hope  that the second opportunity  is realized though the limit depends on the kinematics and we did not find the one where all the partial amplitudes are finite. It might be that the $D=8$ and
$D=10$ cases are better from this point of view. This is still to be found.

The $D=6$ SYM theory sometimes serves as a toy model for gravity since it has a dimensional coupling
and the UV behavior similar to $D=4$ gravity. The results of our analysis show that the finiteness of a theory which is our main goal is not reached so far. 
At the same time, it looks as if the loop by loop cancellation program does not work. It seems that the leading poles do not reveal any additional hidden symmetry, the dual conformal invariance being exploited already. 
Hopefully, the obtained new recursive  relations and the differential equations for the infinite sum of PT open the promising opportunity for the analysis of the UV divergences.

\section{Appendix A: The Master Integrals up to 5 loops}
\begin{figure}[h!]
  \centering
    \includegraphics[scale=0.32]{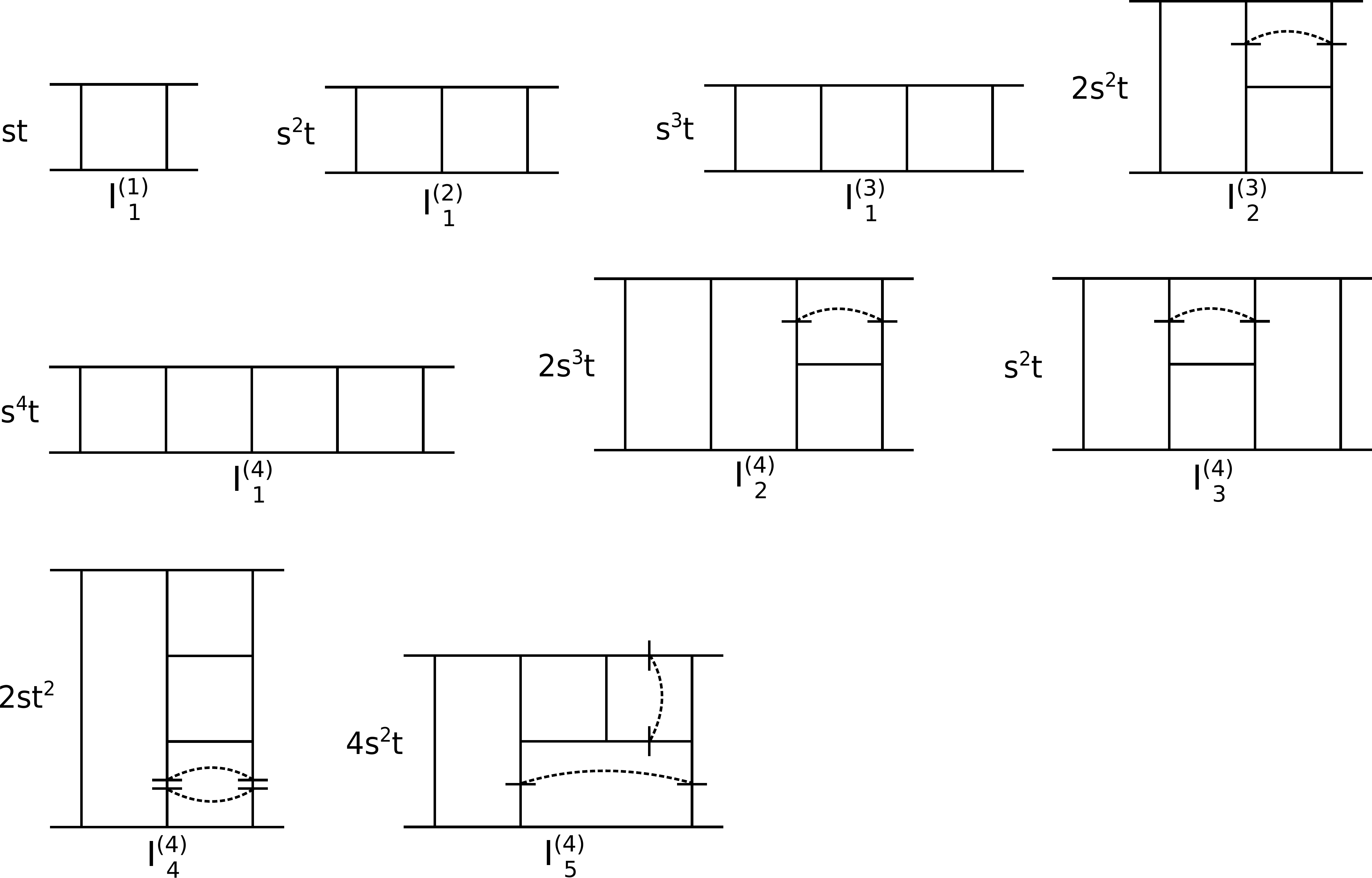}
  \caption{All the diagrams containing the leading pole from one to four loops}\label{diagrams1-4}
\end{figure}

\begin{figure}[!h]
   \centering
  \includegraphics[scale=0.32]{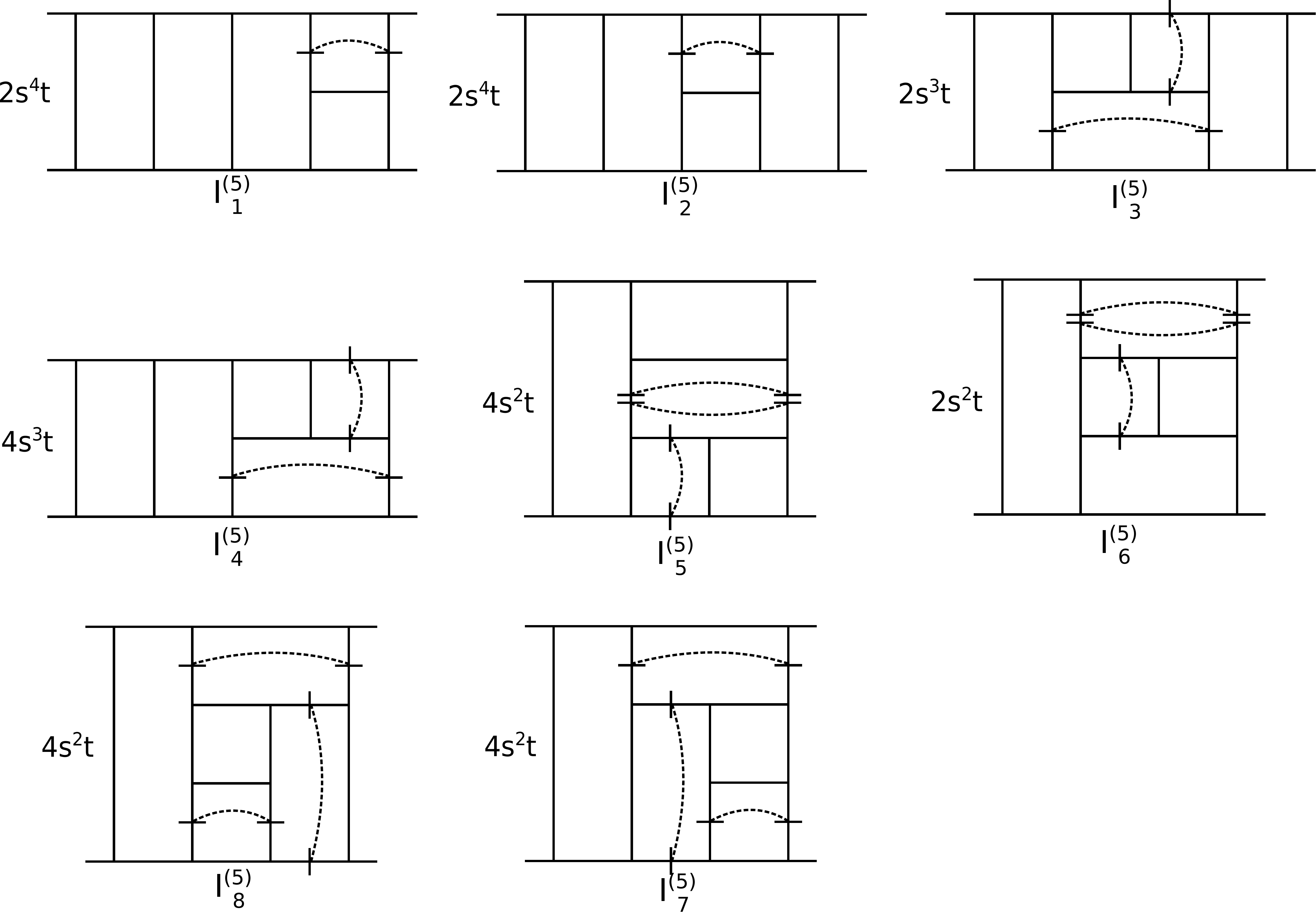}
  \caption{The 5-loop diagrams containing the leading pole in D=6}\label{diagrams5}
\end{figure}

All the diagrams up to 5-loops, which contain the leading pole, are presented in Fig.\ref{diagrams1-4} and Fig.\ref{diagrams5}. The diagrams which can be obtained by the exchange s$\leftrightarrow$t are not shown. The rest of the diagrams  can be found in \cite{DixonBDS45loops}. The combinatorial factor is given on the left side of the diagram. 

\vspace{1cm}

{\bf Acknowledgments}

Financial support from RFBR grant \#  14-02-00494 is kindly acknowledged.


\end{document}